\documentclass[%
 reprint,
 amsmath,amssymb,
 aps,
]{revtex4-2}

\usepackage{siunitx}
\usepackage[super]{nth}
\usepackage{textcomp}
\usepackage{gensymb}
\usepackage{tabularx}
\usepackage{subcaption}
\usepackage{graphicx}
\usepackage{amsmath}
\usepackage{url}

\begin{document}

\preprint{APS/123-QED}

\title{Design of the AWAKE Run 2c transfer lines using numerical optimizers}

\author{R. Ramjiawan}  \altaffiliation[Also at ]{John Adams Institute for Accelerator Science at University of Oxford, Oxford, United Kingdom}\email{rebecca.louise.ramjiawan@cern.ch}

\author{V. Bencini}\altaffiliation[Also at ]{John Adams Institute for Accelerator Science at University of Oxford, Oxford, United Kingdom}
\author{S. D\"{o}bert}
\author{J. Farmer}
\author{E. Gschwendtner}
\author{F.~M. Velotti}
\author{L. Verra}
\author{G. Zevi Della Porta}

\affiliation{CERN, Geneva, Switzerland}

\author{P.~N. Burrows}
\affiliation{John Adams Institute for Accelerator Science at University of Oxford, Oxford, United Kingdom}

\date{\today}

\begin{abstract}

The Advanced Wakefield (AWAKE) Experiment is a proof-of-principle experiment demonstrating the acceleration of electron beams via proton-driven plasma wakefield acceleration. AWAKE Run 1 achieved acceleration of electron beams to \SI{2}{\GeV} and the intention for Run 2 is to build on these results by achieving acceleration to $\sim$\SI{10}{\GeV} with a higher beam quality. As part of the upgrade to Run 2, the existing proton and electron beamlines will be adapted and a second plasma cell and new \SI{150}{\MeV} electron beamline will be added. This new beamline will be required to inject electron bunches with micron-level beam size and stability into the second plasma cell from within the \SI{1}{\m} gap between the two plasma cells. In this paper we describe the techniques used (e.g. numerical optimizers and genetic algorithms) to produce the design of the \SI{150}{\MeV} electron line in order to meet the challenging experimental specifications. Operational techniques are also studied for both electron transfer lines including steering and alignment methods utilising numerical optimizers and beam measurement techniques employing neural networks.
\end{abstract}

\maketitle


\section{Introduction}

\subsection{AWAKE Run 1}

The AWAKE Run 1 experiment at CERN demonstrated that electron beams could be accelerated to GeV energies using plasma wakefield acceleration driven by self-modulated \SI{400}{\GeV} proton bunches~\cite{adli2018acceleration, gschwendtner2016awake}. The plasma was produced via the ionization of Rubidium in a \num{10}-m-long vapor cell with a high-power laser pulse, forming a \num{1}-mm-radius plasma channel~\cite{oz2014novel}. The wakefield driver consisted of a \SI{400}{\GeV}, \num{12}-cm-long proton beam from the CERN Super Proton Synchrotron (SPS) which was injected with a beam size of \SI{200}{\um} into the plasma where it underwent self-modulation into a train of microbunches~\cite{adli2019experimental,turner2019experimental}. The microbunches had lengths approximately equal to half of the plasma wavelength and together resonantly drove large-amplitude plasma wakefields. The laser pulse used to ionize the plasma co-propagated in the plasma cell with the proton driver-beam and the self-modulation of the proton-bunch behind the laser pulse was seeded by the ionization front of the laser pulse. In this way, the the proton bunch behind the laser pulse underwent the phase-reproducible Seeded Self-Modulation (SSM)~\cite{batsch2021transition}. 

With the nominal plasma electron density of $7\times10^{14}~\SI{}{cm^{-3}}$, the maximum accelerating gradient was of order GV/m~\cite{lotov2014parameter}. To probe the accelerating gradients of the wakefields, \SI{18.84}{\MeV} witness electron bunches were injected into the wakes~\cite{adli2018acceleration}. The electron beamline comprised an S-band, RF photo-cathode gun producing electron bunches which were accelerated with a traveling-wave booster structure to 16-\SI{20}{\MeV}. A transfer line~\cite{schmidt2015awake} transported and focused these beams so that they could be injected on-axis into the plasma cell with a beam size of \SI{250}{\um}~\cite{kim2020commissioning}. Electrons trapped in the focusing, accelerating phase of the wakefield were accelerated to \SI{2.0\pm0.1}{\GeV}~\cite{adli2018acceleration}, as measured with a magnetic spectrometer.

\subsection{AWAKE Run 2}
The goal of AWAKE Run 2 is to achieve acceleration to higher energies while maintaining a smaller emittance and energy spread than Run 1~\cite{muggli2020physics}. Towards this objective, Run 2 will be split into four intermediary stages~\cite{edda2021awake}:
\begin{itemize}
    \item \textbf{2a:} demonstrate the seeding of the self-modulation of the full proton bunch with an electron beam to ensure modulation of the whole bunch is phase reproducible and stable, 
    \item \textbf{2b:} introduce a density step in the plasma to stabilise the proton bunch self-modulation~\cite{caldwell2011plasma},
    \item \textbf{2c:} separate the proton bunch self-modulation and the electron bunch acceleration into separate plasma cells to isolate the defocusing fields of the unmodulated proton bunch from the electron bunch~\cite{gorn2018response},
    \item \textbf{2d:} demonstrate the scalability of the experiment to longer plasma cells and higher energies, where such an accelerator could have applications for an electron-proton collider or for fixed target experiments. 
\end{itemize}

\begin{figure*}
    \includegraphics[width=1\linewidth]{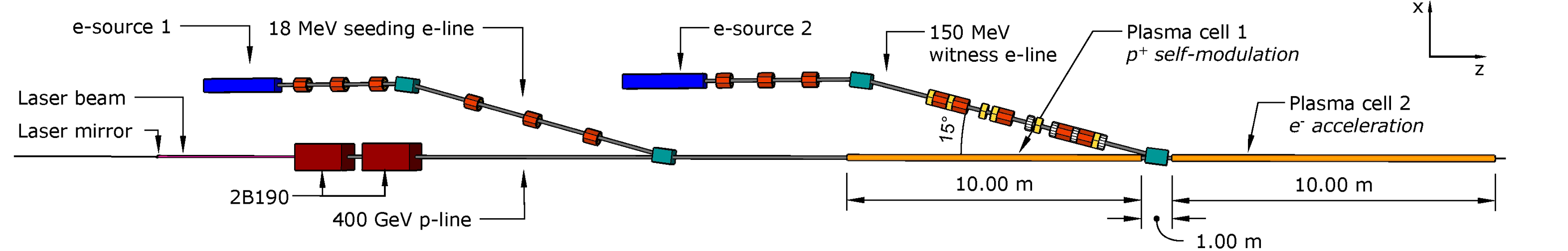}
    \caption{Schematic of the configuration of the two electron beamlines, plasma cells and a section of the proton transfer line. Dipoles are shown in cyan, the quadrupoles in red, the sextupoles in yellow and the octupoles in white.}
    \label{fig:SchematicLines}
\end{figure*}

Run 2a will study the electron-seeding of the proton bunch self modulation and Run 2b will investigate the use of a density step to stabilise the self modulation process; both of these features will then be incorporated into Run 2c. 

In this paper we discuss the studies towards the baseline design of the Run 2c transfer line needed to inject witness electron bunches into the second plasma cell to probe the accelerating gradients of the wakefields. A schematic of the proposed Run 2c beamline layout is shown in Fig.~\ref{fig:SchematicLines} showing the configuration of the proton and electron beamlines. Several changes will be needed to adapt the Run 1 experimental set up for Run 2c. To incorporate the additional seeding electron transfer line, the Run 1 plasma cell is to be moved \SI{40}{\m} downstream, requiring the reconfiguration of the proton beamline. To minimise the defocusing of the proton beam between the two plasma cells the gap should be $<\SI{1}{\m}$~\cite{adli2016towards}, and ideally as short as possible, constraining the footprint of the witness transfer line. To achieve both a small energy spread and emittance conservation throughout acceleration, the injected electron beam parameters must be carefully chosen; this is discussed in Section~\ref{TLSpec}~\cite{olsen2018emittance}. The witness beam parameters for Run 2c compared with Run 1 are presented in Table~\ref{table:Run1Run2}.

The seeding electron-beamline will inject $\sim$\SI{18}{\MeV} bunches into the first plasma cell. The parameters for this line will be determined as a result of the Run 2a studies and it is foreseen to be adapted from the Run 1 electron beamline. In Section~\ref{NeuralNetworks} we present a proposal for a method which could be used to estimate the relative alignment of the Run 2c proton and seeding electron-beams using neural networks.

\begin{table}[htbp]
\caption{Beam parameters of the witness electron transfer lines for AWAKE Runs 1 and 2c~\cite{adli2018acceleration,pepitone2018electron}.}
\newcolumntype{Y}{>{\centering\arraybackslash}X}
\newcolumntype{M}{>{\centering\arraybackslash}m{1.7cm}}
\begin{center}
\begin{tabularx}{\linewidth}{X M M M}\toprule\\[-0.9em]
\textbf{Parameter} & \textbf{Unit}  & \textbf{Run 1} & \textbf{Run 2c}\\\hline\\[-0.9em]
Beam energy & [MeV] & 18.84 & 150\\
Charge& [pC] & 656 &100\\
Bunch length &[fs] & 4000 & 200 \\
Energy spread &[\%] & 0.5 & 0.2\\
Norm. emittance& [mm~mrad] & 11-14 & 2\\
\toprule
\end{tabularx}
\end{center}
\label{table:Run1Run2}
\end{table}

\subsection{The use of optimizers in beamline design and operation}

Numerical optimizers are powerful tools for beamline design; see~\cite{taheri2019genetic,ROSSETTICONTI201884} for examples of existing studies exploring their use for electron beamline design and optimization.

Here we present proposals for the design of the Run 2c witness electron transfer line, alongside a discussion of the use of numerical optimizers during the design process. In Sections~\ref{BeamSteering} and~\ref{NeuralNetworks} we discuss also the operational challenges expected for the seeding and witness transfer lines and highlight where machine learning or optimization techniques could be exploited.

\section{Transfer line design}

\subsection{Witness electron transfer line specifications}
\label{TLSpec}

The specifications for beam parameters at the injection point derive from the need for the witness beam to be `matched' to the plasma to mitigate transverse betatron oscillations of the beam envelope propagating in the plasma which would cause beam emittance growth. For the electron beam to be matched to the plasma, the beam size should satisfy
\begin{equation}
    \sigma^{4}=\frac{2\varepsilon_0m_ec^2\gamma}{n_{pe}e^2}\epsilon^2,
\end{equation}
where the Lorentz $\gamma = 293.5$, $m_e$ is the mass of an electron, $c$ is the speed of light, $\varepsilon_0$ is the vacuum permittivity, $e$ is the electron charge, the normalised emittance $\epsilon = 2$~mm~mrad and the plasma density $n_{pe}$ has baseline values: $2\times10^{14}$ \SI{}{cm^{-3}} or $7\times10^{14}$ \SI{}{cm^{-3}}. For beam energy of 150 MeV with the higher plasma density this would correspond to a matched beam size of
\begin{equation}
\sigma^*=\SI{5.75}{\um}.  
\end{equation}
Further specifications for the beam at the injection-point are given in Table~\ref{table:ESpec}.

The injected witness bunch should have a length of $\sim$\SI{60}{\um}, a specification deriving from the need to be within the regime of optimal beam loading so that a small energy spread is conserved during acceleration~\cite{olsen2018emittance, pepitone2018electron}. To minimize emittance growth throughout acceleration, there should be sufficient charge density in the witness bunch to be able to drive a full blow-out of the electrons remaining in the plasma wakefield `bubble'~\cite{olsen2018emittance}. The emittance growth during electron acceleration increases quickly with the transverse relative offset between the proton and electron beam. Simulations of the witness beam propagation in the plasma have shown that, for a nominal \SI{2}{\mm}~mrad emittance beam, to maintain an acceptable beam quality the relative beam offset should not exceed \SI{13}{\um} and the beam size should not exceed 150\% of the nominal value.

\begin{table}[htbp]
\caption{Specification for the bunch parameters at the injection-point of the AWAKE Run 2c witness transfer line.}
\newcolumntype{Y}{>{\centering\arraybackslash}X}
\newcolumntype{M}{>{\centering\arraybackslash}m{2cm}}
\begin{center}
\begin{tabularx}{0.7\linewidth}{Y Y }\toprule\\[-0.9em]
\textbf{Parameter} & \textbf{Specification}\\ \hline\\[-0.9em]
$\beta_{x,y}$ & $\SI{4.87}{\mm}$\\
$\alpha_{x,y}$ & $0.0$\\
$D_{x,y}$  &$\SI{0}{\m}$\\
$\sigma_{x,y}$ &$\SI{5.75}{\um} $\\
$\sigma_{z}$ &$\SI{60}{\um} $\\
$\epsilon_{x,y}$ &$\SI{2}{\mm}$~mrad\\
\toprule
\end{tabularx}
\end{center}
\label{table:ESpec}
\end{table}

The footprint of the witness beamline is constrained by the placement of the two plasma cells, the limited tunnel width and the location of the seeding electron beamline. This constrains the  width of the beamline to $<\SI{3}{\m}$, and the length to $<\SI{25}{\m}$; a two-dipole dog-leg design was selected to satisfy these restrictions. The dimensions of the dog-leg are determined by the position and bending angle of the dipoles. 

A 15\degree~bending angle was selected as this was large enough that the beam-pipe would not intersect with the plasma cell but not so high that the beamline exceeded the tunnel width. For a two-dipole achromatic dog-leg, the first-order isochronous parameter, $R_{56}$, cannot be compensated, so that the transfer line would not be both achromatic and isochronous. To meet the bunch length specification of $\sigma_z=\SI{60}{\um}$ at the plasma injection-point, it is proposed that the line have a shortening effect on the bunch, counteracted by injecting a correspondingly longer bunch into the transfer line. For the transfer line to have a shortening effect on the bunch there must be a positive energy-longitudinal correlation, which  based on simulations of the electron injector is expected to be feasible.

\subsection{Transfer line simulations}

\begin{figure}[htbp]
\centering
\includegraphics[width=0.99\linewidth]{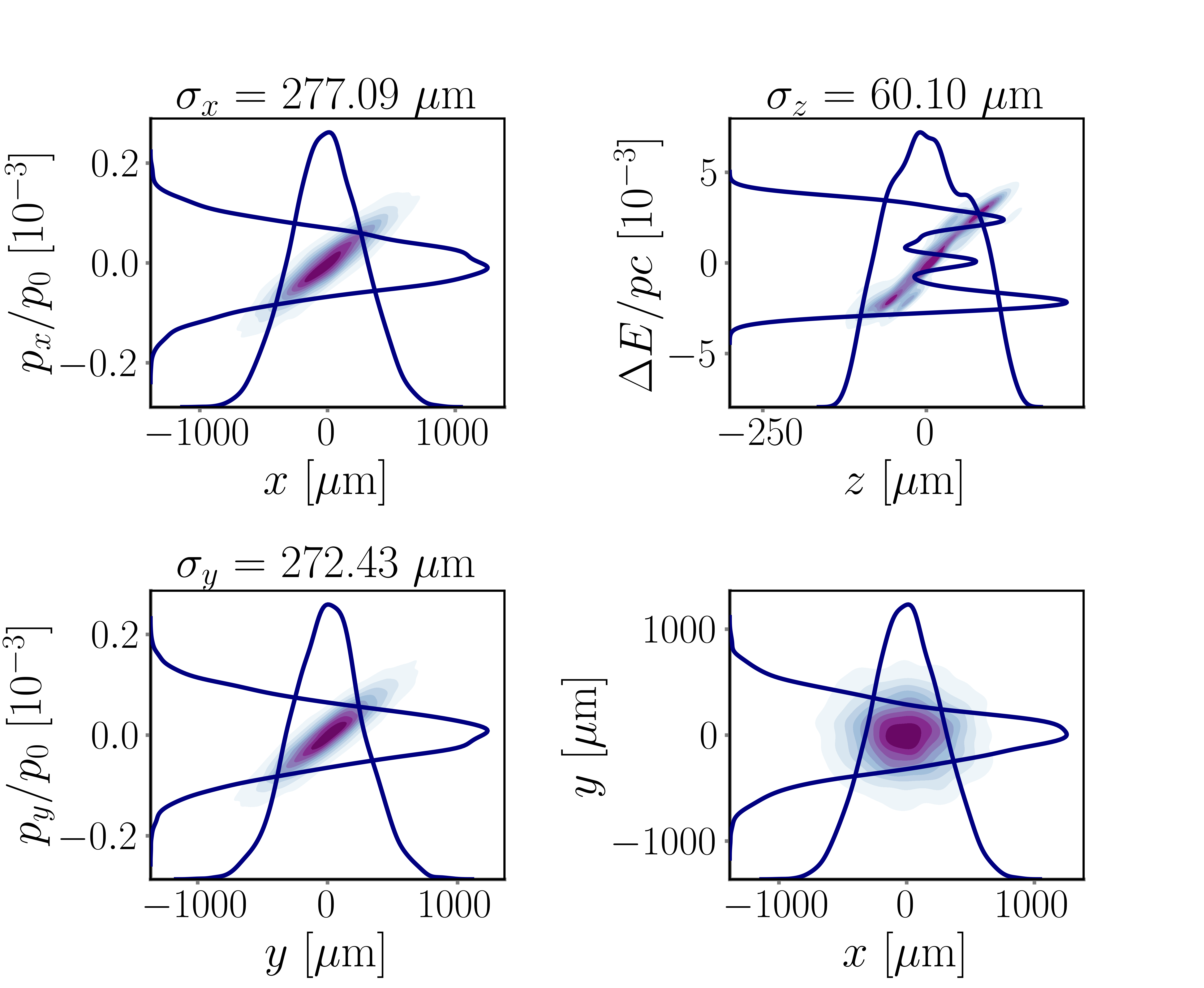}
\caption{Generated input beam distribution of \num{100000} macro-particles with $\beta_{x,y}=\SI{11}{\m}$, $\alpha_{x,y}=-2.1$ and $\epsilon_{x,y}=\SI{2}{\mm}$~mrad. The $E$-$z$ distribution was taken from simulations of the electron injector and scaled to have the nominal bunch length and energy spread.
}
\label{fig:input_distr}
\end{figure}

The simulation code MAD-X~\cite{madx} was used to model the beam transport, with the bunch tracking simulated using a MAD-X implementation of PTC (Polymorphic Tracking Code)~\cite{forest2002introduction}. The non-linear effects were challenging for this design, so six dimensional particle tracking was essential for modelling the behaviour of the line. 

Simulations of the electron injector were used to produce an input bunch, shown in Fig.~\ref{fig:input_distr}, which was tracked through the transfer line to calculate the beam parameters at the injection-point. The input bunch had \num{100000} macro-particles and was designed such that the distributions in the $x$, $y$, $px$ and $py$ planes were Gaussian distributions cut at $3\sigma$, with standard deviations matching simulations of the beam from the electron injector. The $z$-$pz$ distribution was taken from the electron injector tracking simulation and scaled to match the nominal Run 2 bunch length and energy spread - thus preserving the simulated $E$-$z$ correlation from the electron gun.

\subsection{Numerical optimization for transfer line design}

Optimization is the problem of finding a set of inputs for a function which corresponds to a maximum or minimum of that function. For an optimization problem, the function to be minimized or maximized is called the `objective function'. An optimization algorithm specifies the method used to iteratively choose inputs, evaluate the objective function and compare solutions with the aim of moving towards the optimal solution. If multiple parameters need to be optimized, a single objective function can be formed as a weighted sum of the individual objectives (`scalarization') or they can form multiple objective functions (Multi-Objective Optimization). With scalarization, the weights should be tuned to achieve the best performance from the optimization algorithm. For the design of the witness transfer line, both single- and mutli-objective optimization was tested. The objective functions were adjusted at every step of the design process to take into account the current status of the design. For example, as the design progressed, higher order parameters became increasingly important and their weightings in the objective function were increased. 

The objective functions were formed as a weighted Mean Squared Error (MSE),
\begin{equation}
\frac{1}{n}\sum_{t=1}^{n}w_i(y_i - y_i^{target})^2,
\end{equation}
where $w_i$ are weights, and $y_i$ and $y_i^{target}$ are the simulated and target parameters respectively. Depending on the progress of the design, $y_{i,..,n}$ may include parameters such as the beam size, dispersion or Twiss parameters. The input variables and target parameters were each normalized by their respective maximum values. 

Simulations of the electron propagation in plasma have shown that the beam distribution at the injection-point is an important parameter for maintaining a high beam quality during acceleration. In order to optimize the beam distribution, the Kullback–Leibler (KL) divergence~\cite{kullback1951information} was used to quantify the difference between the tracked bunch distribution and an ideal distribution. The K-L divergence is defined as
\begin{equation}
D_{KL}(p \;\vert \vert\; q) = \sum^n_{i=1} p(x_i)\log(\frac{p(x_i)}{q(x_i)}),
\end{equation}
where $q(x)$ is the distribution under test and $p(x)$ the true distribution. The lower the KL-divergence, the better the agreement between the two distributions. This was particularly useful for producing transfer line designs where the injected bunch needed to be Gaussian in six dimensions ($x$, $px$, $y$, $py$, $z$, $pt$). Alternatively, if, rather than a Gaussian beam, a higher charge density is needed in the bunch core, then this corresponds to a high-kurtosis design, where the kurtosis is a measure of the tailed-ness of the distribution, defined as $\mathrm{Kurt} ={\mu_4}/{\sigma^4}$, where $\mu_4$ is the fourth central moment and $\sigma$ is the standard deviation. In this case, the kurtosis was a helpful parameter to include in the objective function. For the AWAKE Run 2c experiment, the effect of the beam distribution on the quality of the acceleration is important to study and, so, being able to optimize the optics to achieve a given distribution or kurtosis would be helpful.

For the initial design stages, the simulations were performed with fewer macro-particles, sacrificing accuracy for speed. Once a coarse solution was found, fine-tuning was performed with a larger number of macro-particles. An essential parameter to include in the objective function was the fraction of the macro-particles lost throughout the transfer line due to aperture constraints. When optimizing magnet positions, it was also important to heavily penalise the overlap of magnets in the model.

\subsection{Optimization algorithms}

This choice of the optimization target, and the way it is calculated, makes it a black-box-like optimization problem. This thus constrains the choice of algorithms to be used to derivative-free optimizers, random search algorithms and genetic algorithms. 

A range of optimization algorithms were used throughout the design process for the Run 2c witness transfer line. When selecting an optimization algorithm, considerations may include whether there are constraints on the input variables or objective functions, the cost of each function evaluation and whether the function is convex. A brief overview of the main optimization algorithms used for this design are given below. 

\paragraph{Genetic Algorithm}
A genetic algorithm (GA) is an optimization method based on a natural selection process akin to biological evolution. The population is a group of individual solutions and at each step, the GA randomly selects individuals from the population and uses them as `parents' to produce the `children' for the next generation. GAs use three basic operators, selection, crossover (mating), and mutation; over successive generations, the population evolves toward an optimal solution~\cite{mitchell1998introduction, nsgaii}. 

\paragraph{Multi-Objective Optimization}

Multi-objective optimization problems seek to optimize two or more objective functions simultaneously. If there is not a single solution which simultaneously optimises all objectives, a set of optimal solutions called a `Pareto set' may be found. The Pareto set are a set of solutions which are not dominated by any other solutions, where solution $A$ is dominant if and only if there is no objective of $A$ worse than that objective of $B$ and there is at least one objective of $A$ strictly better than that objective of $B$.

Non-dominated Sorting Genetic Algorithm II (NSGA-II)~\cite{nsgaii} is a multi-objective optimization algorithm using GAs, exploiting the concept of dominant and non-dominant solutions to help quantify the fitness of solutions in a population. The algorithm selects solutions for parents using a fitness metric based on `non-dominant sorting' and `crowd distance sorting' of solutions with the goal of obtaining a strong and diverse set of parents.

\paragraph{Powell}
A gradient-free, unconstrained optimization algorithm which minimizes a function by using sequential line searches along search vectors, often the axes of the input variables~\cite{powell1970new}. After each line search, the algorithm moves to the minimum found before progressing to the next vector. The process continues until no direction can be found that will decrease the function.

\paragraph{Nelder-Mead}

The Nelder-Mead algorithm is a search method to find the optimum value of an objective function in an $n$-dimensional space by using a simplex shape to explore the domain~\cite{nelder1965simplex}. At each iteration, one vertex of the simplex moves towards a more optimal solution, where for each step several possible adjustments may be tested before one is selected. 

\paragraph{Py-BOBYQA}
Py-BOBYQA~\cite{numericalalgorithmsgroup} is Python implementation of the Bound Optimization BY Quadratic Approximation (BOBYQA) algorithm~\cite{powell2009bobyqa} which employs a `black-box' approach to solving the minimization of function $F(x)$ by approximating it by a quadratic function at each iteration. It seeks to minimise $F(x)$ while respecting the bounds $a_i$ and $b_i$,
\begin{equation}
    a_i \leq x_i \leq b_i,\;\;\;i=1,2,..,n.
\end{equation}
BOBYQA is particularly useful for problems where evaluations of the objective function are time consuming, computationally intensive or costly.

\section{Design of the Run 2c witness transfer line}

The full design of the witness transfer line was created over several intermediary stages, with the optimization algorithms and objective function adapted for each specific stage. To begin with, a transfer line design was produced from only dipoles and quadrupoles. The dipole position and bending angle was fixed by the desired geometry of the line and the quadrupole strengths and positions were the variables to be optimized. As the design for the transfer line progressed, non-linear effects proved significant, requiring the addition of sextupoles and octupoles. These were added sequentially as required and the transfer line was iteratively re-optimized until a design was achieved which met the experimental specifications.

\subsection{Optimization of quadrupole positions and strengths}
\label{section:quads}

Typically for a transfer line with a very small focal point, the distances between the final focusing magnets and the focal point are minimised so as to reduce chromatic contributions to the beam size. Due to the proximity of the transfer line to the plasma cell (Fig.~\ref{fig:SchematicLines}), to prevent the intersection of the final quadrupole with the plasma cell, the distance between the quadrupole and the focal point must be at least \SI{1.9}{\m}. 

An initial transfer line design was constructed with the two dipoles, a quadrupole triplet before the dog-leg and five quadrupoles within the dog-leg. The triplet was intended to focus the beam to a waist before the dog-leg, so that the dog-leg can transport the waist to the merging point. To quantify the performance of the transfer line, a 6D beam distribution (Fig.~\ref{fig:input_distr}) was tracked through the beamline to the injection-point. The input variables for optimization were the positions and strengths of the eight quadrupoles. Within the dog-leg, the symmetric pairs of quadrupoles have strengths of equal magnitude so that there are 14 dimensions to the optimization problem. 

Initially, the tracked beam distribution had only \num{5000} macro-particles which was increased with each successive stage. To produce a coarse initial design, only a reduced set of parameters were included in the objective function to prioritise the most challenging and important parameters for the overall design. The objective function, Eq.~\ref{eq:quad_obj} included the horizontal and vertical beam sizes ($\sigma_x$, $\sigma_y$), dispersion ($D_x$, $D_y$) at the merging point, and the number of particles lost ($N_\mathrm{loss}$). The number of particles lost was heavily weighted to penalise the algorithm if the beam envelope exceeded the apertures. Any overlap between magnets was penalised with a large weight to the whole objective function, $M$.

\begin{equation}
\begin{split}
f_0(x) = M\log\big\{ & w_1(\sigma_x - \sigma_x^\mathrm{target})^2 + w_2(\sigma_y - \sigma_y^\mathrm{target})^2 \\
+&w_3(D_x - D_x^\mathrm{target})^2 +
w_4(D_y - D_y^\mathrm{target})^2 \\
+&w_5N_\mathrm{loss}\big\}
\end{split}
\label{eq:quad_obj}
\end{equation}

\begin{figure}[htbp]
\centering
\includegraphics[width=0.99\linewidth]{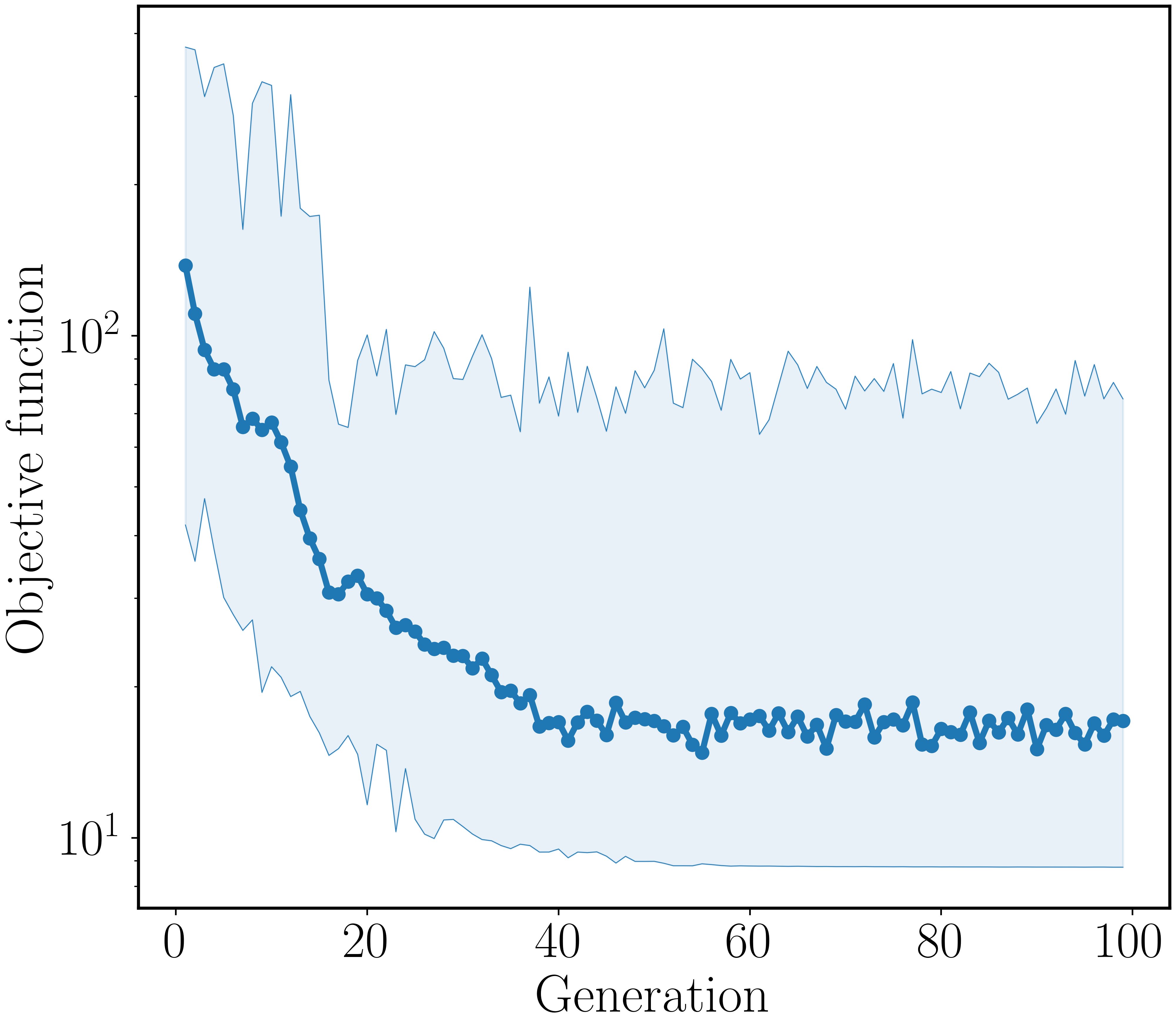}
\caption{Mean (data-points) and range (shaded region) of the objective function evaluations from a population of solutions used by a Genetic Algorithm for the optimization of quadrupole strengths and positions.}
\label{fig:objective_quads}
\end{figure}

\begin{figure}[htbp]
\centering
\includegraphics[width=0.99\linewidth]{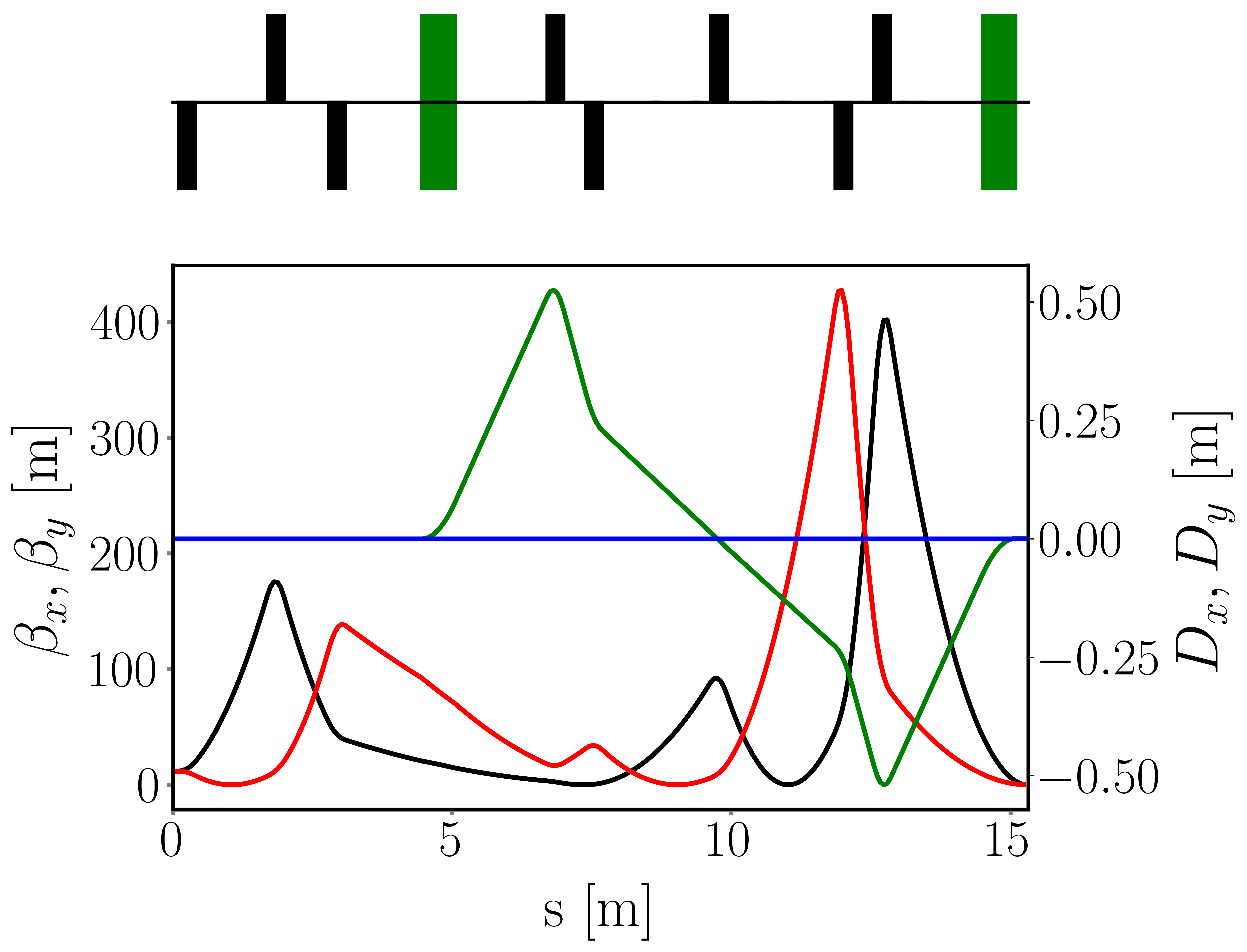}
\caption{MAD-X simulation of the transfer line design selected via the optimization shown in Fig.~\ref{fig:objective_quads}. Twiss parameters $\beta_x$ (black), $\beta_y$ (red) and dispersion $D_x$ (green) and $D_y$ (blue) are shown below, with a synoptic overview of the transfer line above, with dipoles (green) and quadrupoles (black).}
\label{fig:twiss_quads}
\end{figure}

\begin{figure}[htbp]
\centering
\includegraphics[width=0.99\linewidth]{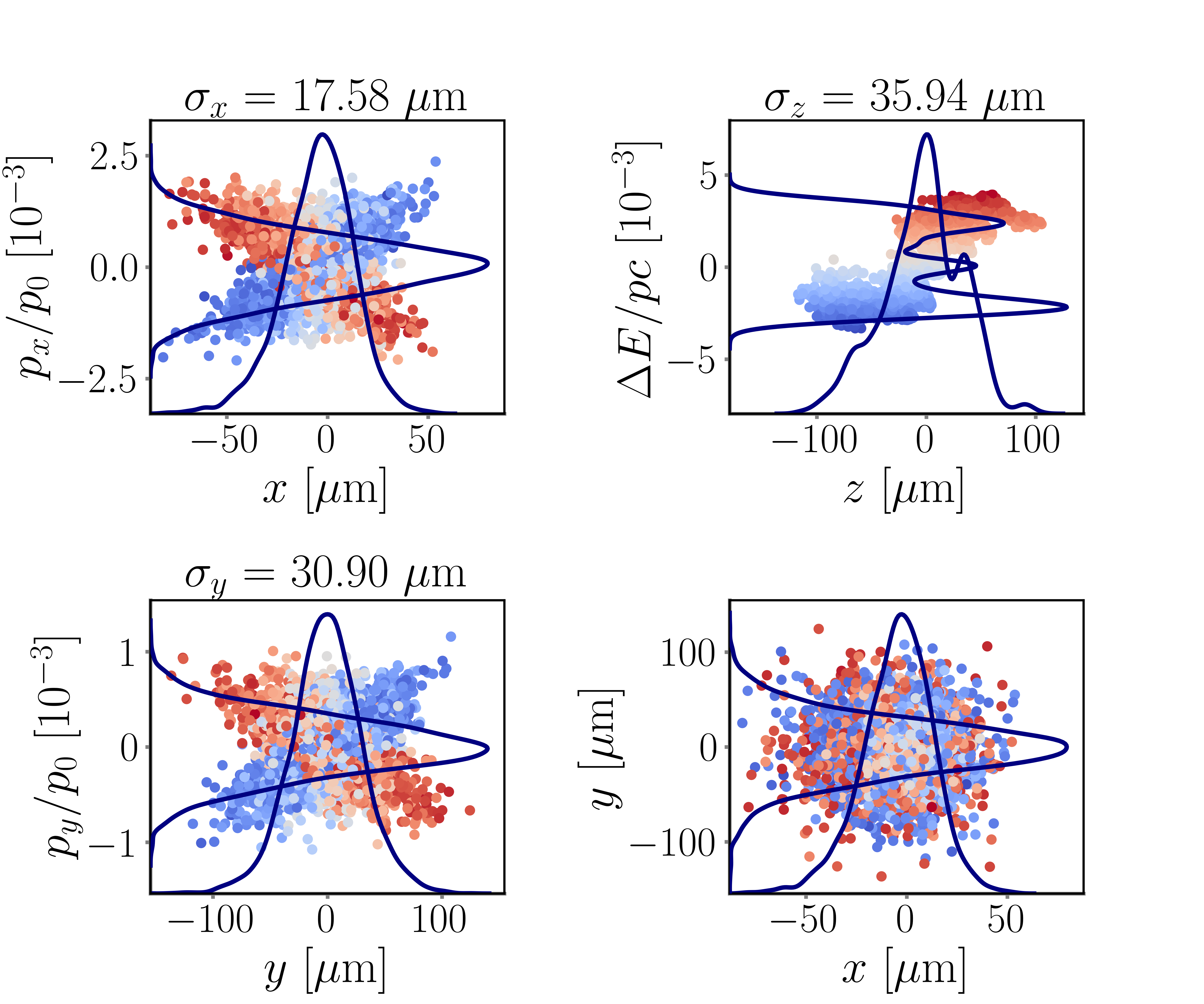}
\caption{2D projections of the distribution of a bunch tracked through the transfer line shown in Fig.~\ref{fig:twiss_quads} to the plasma injection-point. The color of the data points denote the momentum offset of the macro-particles.}
\label{fig:plotmat_quads}
\end{figure}

As a first step, a Genetic Algorithm was tested to try to find a global minimum. A population size of \num{200} with \num{300} offspring per generation were used. The mutation parameter was set high to encourage the exploration of the parameter space and to prevent the algorithm from getting trapped in local minima. The evolution of the mean objective function with successive generations of the genetic algorithm is presented in Fig.~\ref{fig:objective_quads} where the minimum objective function in the generation decreased over $\sim$\num{60} generations before stabilising. The solution corresponding to the minimum objective function evaluation had horizontal and vertical bunch sizes at injection of \SI{17.58}{\um} and \SI{30.90}{\um} respectively. The optics corresponding to this solution are shown in Fig.~\ref{fig:twiss_quads} and the bunch distributions at the injection-point are presented in Fig.~\ref{fig:plotmat_quads}. The macro-particles are colour-coded by their $\Delta E$ values, highlighting the chromatic contributions to the beam size. In order to minimize the beam size, this effect was mitigated by the addition of sextupoles to the design.

\begin{figure}[htbp]
\centering
\includegraphics[width=0.95\linewidth]{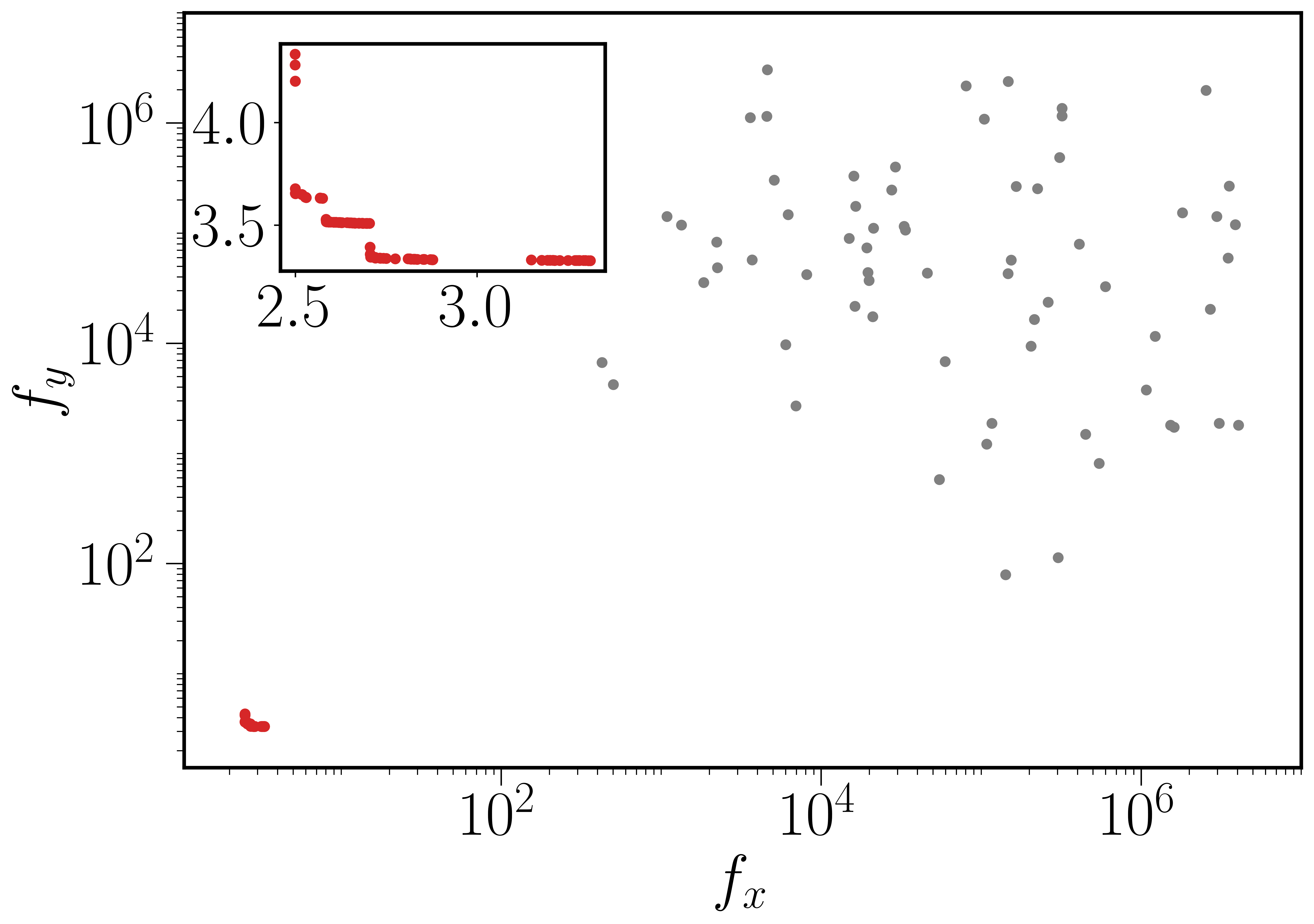}
\caption{Points on the pareto front at the final evaluation (red) and the population from the first evalution (gray) of the Multi-Objective Optimization of quadrupole strengths and positions; the axes are the two objective functions constructed from the horizontal and vertical components of Eq.~\ref{eq:sext_obj}. The inset shows an expanded view of the final pareto front. Only viable solutions with no magnet overlap are included.}
\label{fig:MOO}
\end{figure}

Results from testing a Genetic Algorithm for optimizing this transfer line design suggested that the horizontal and vertical beam sizes could each be reduced but at the expense of the other plane. To investigate the limits for minimizing the beam sizes, the multi-objective optimization algorithm, NSGA-II, was used to optimize the horizontal and vertical parameters in Eq.~\ref{eq:quad_obj} separately. Fig.~\ref{fig:MOO} shows the results from \num{25000} iterations of the algorithm with a population of 100, after which the final Pareto front was found to be in good agreement with the single objective Genetic algorithm results. Fig.~\ref{fig:MOO} shows that even when minimizing the horizontal and vertical beam sizes independently it is not possible to satisfy the experimental requirements. 

\subsection{Addition and optimization of sextupoles and octupoles}
\label{section:sextupoles}

\begin{figure}[htbp]
\centering
\includegraphics[width=0.99\linewidth]{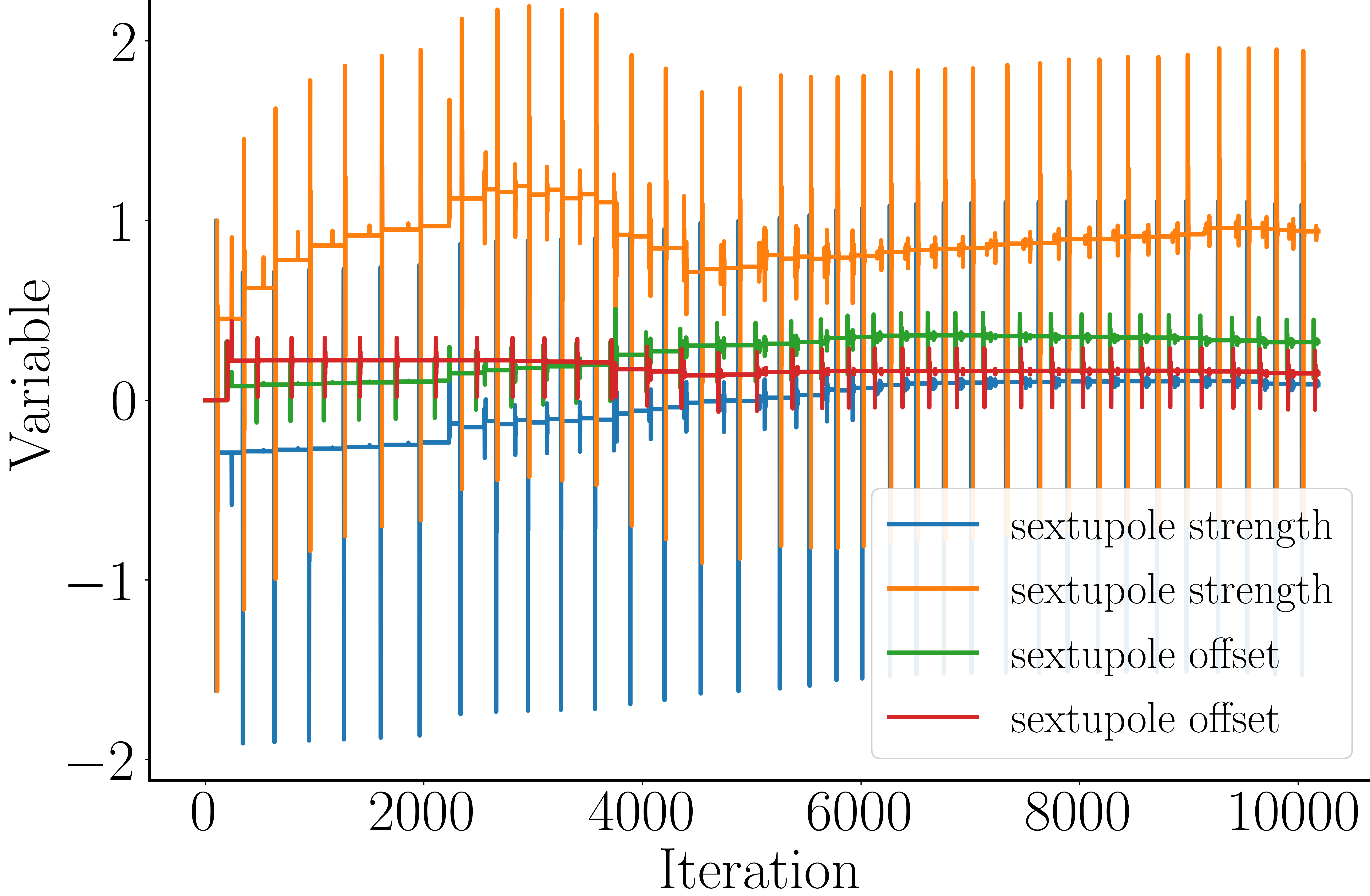}
\caption{Optimization of the quadrupole and sextupole strengths and positions using the Powell algorithm, starting from the design described in Section~\ref{section:quads}. Four of the normalized input variables are shown as a function of the iteration of the Powell optimization algorithm.}
\label{fig:sext_progress}
\end{figure}

To combat the chromatic contributions to the beam size at the injection-point, sextupoles were added, initially chosen to be at regions of high ${D_x}/{x}$ and ${D_y}/{y}$ and at phase advances to have maximum impact at the injection-point. The sextupole positions were then included, along with the strengths of the sextupoles, to the list of inputs for the optimization algorithm. The Powell optimiser proved to perform well for this stage, as it coped well with the high dimenionality and was explorative of the parameter space. This optimization was performed over many iterations and an example of the progress of a few of the input variables for one such iteration is presented in Fig.~\ref{fig:sext_progress}. This shows clearly the methodical approach of the Powell algorithm in scanning each dimension successively. Once an initial optimization of the layout of the transfer line with sextupoles had been performed, finer adjustments were made for which the weights in the objective function were adjusted and higher-order parameters like $\alpha_{x,y}$, $D'_{x,y}$ and the K-L divergence were included or given a higher weighting. This objective function could be written as

\begin{equation}
\begin{split}
f_1(x) = M\log\big\{ & w_1(\sigma_x - \sigma_x^\mathrm{target})^2 + w_2(\sigma_y - \sigma_y^\mathrm{target})^2\\
+&w_3(D_x - D_x^\mathrm{target})^2 +
w_4(D_y - D_y^\mathrm{target})^2\\
+&w_5(\alpha_x - \alpha_x^\mathrm{target})^2 +
w_6(\alpha_y - \alpha_y^\mathrm{target})^2\\
+&w_5N_\mathrm{loss}+w_6K{\text -}L_\mathrm{div.}\big\},
\end{split}
\label{eq:sext_obj}
\end{equation}
where $\alpha_{x,y}$ are Twiss parameters and $K{\text -}L_\mathrm{div.}$ is the K-L divergence.

\begin{figure}[htbp]
\centering
\includegraphics[width=0.99\linewidth]{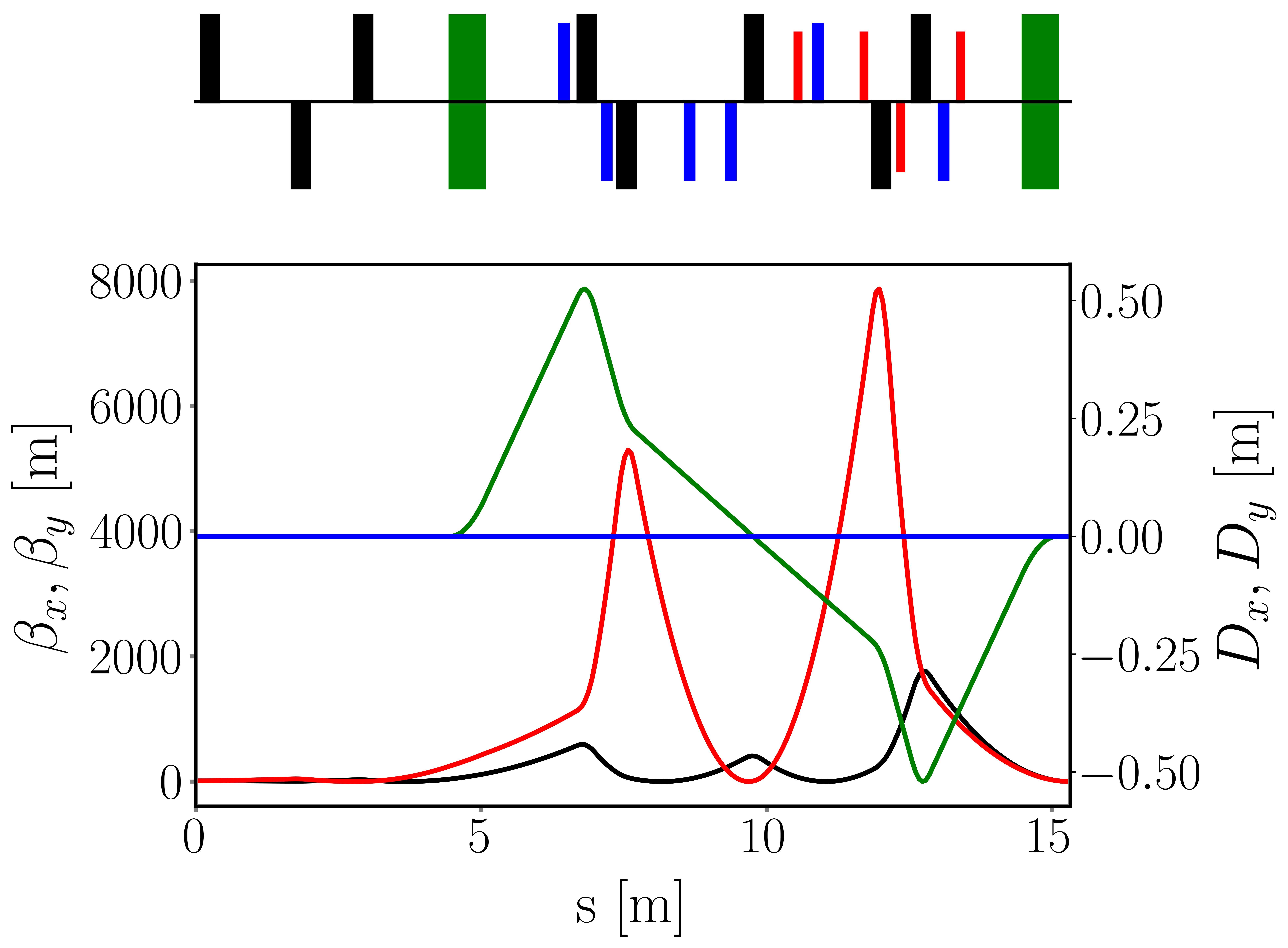}
\caption{MAD-X simulation of the transfer line design after the optimization of the sextupole and octupole positions and strengths. The colors in the plot are as for Fig.~\ref{fig:twiss_quads} with the addition of sextupoles (blue) and octupoles (red) to the transfer line synoptic.}
\label{fig:twiss_params}
\end{figure}

\begin{figure}[htbp]
\centering
\includegraphics[width=0.99\linewidth]{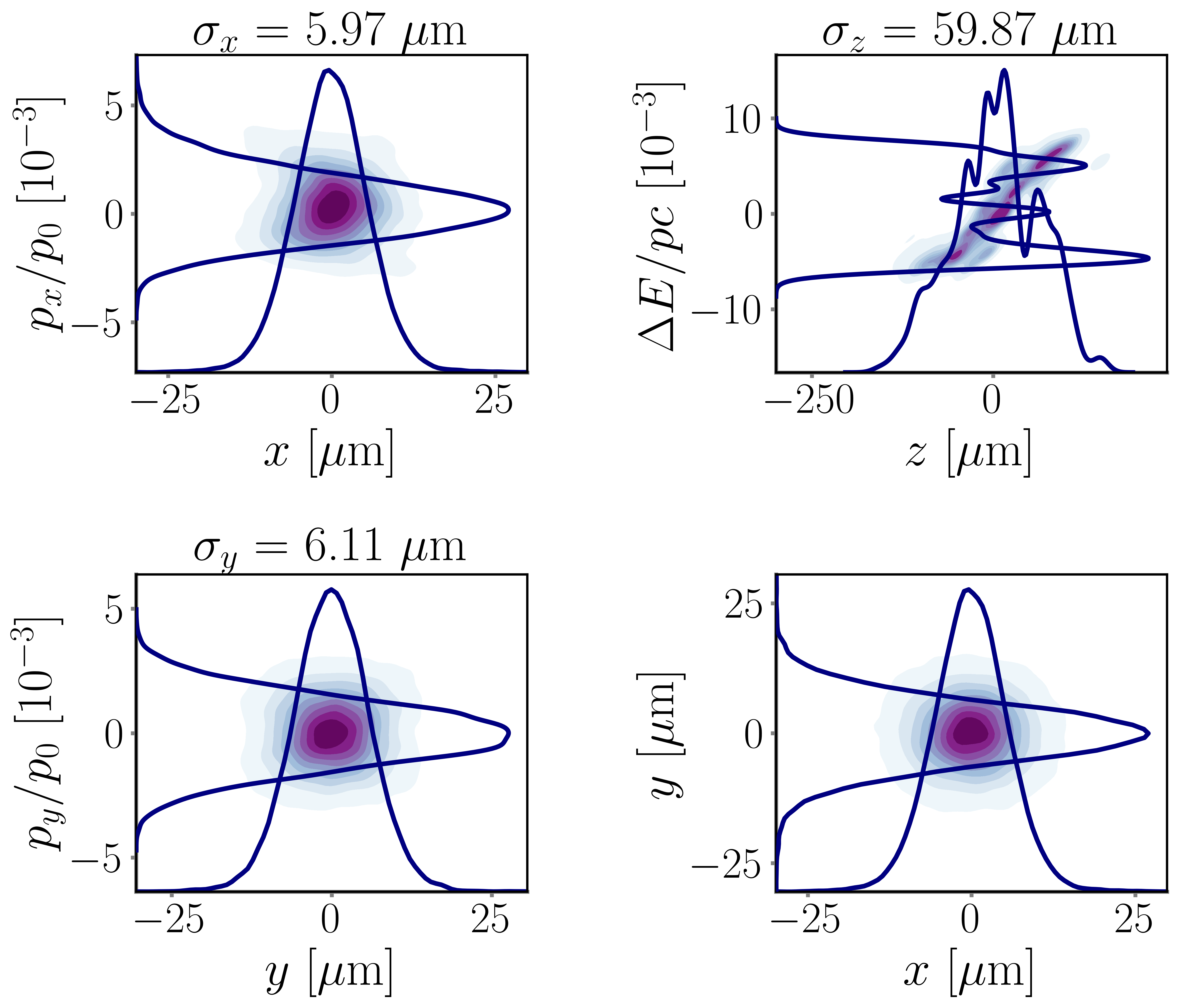}
\caption{2D projections of the nominal beam distribution tracked through the transfer line shown in Fig.~\ref{fig:twiss_params} to the injection-point.}
\label{fig:beam_dist}
\end{figure}

Even after a solution was found for which the sextupoles would mitigate the chromatic effects, there were remaining detuning-with-amplitude effects contributing to the beam size. Without the addition of octupoles to correct for these effects it seemed impossible to reduce the beam size below \SI{8}{\um} either horizontally or vertically. Octupoles were added incrementally and their positions and strengths optimized using the same process as for the sextupoles. Detuning with amplitude effects could not be corrected locally throughout the entire line, so the focus was on reducing these effects at the injection-point. The optics for the resulting transfer line are presented in Fig.~\ref{fig:twiss_params} showing the final placements of the six sextupoles and four octupoles. The corresponding beam distributions at the injection-point are presented in Fig.~\ref{fig:beam_dist} with the beam parameters given in Table~\ref{table:final_params}, showing that this transfer line design meets experimental requirements for the beam size. For the transfer line to meet the nominal bunch length, $\sigma_z=\SI{60}{\um}$, the input bunch length should be 40\% longer.

\begin{table}[htbp]
\caption{Beam size and linear optics parameters at the injection-point for a transfer line shown in Fig.~\ref{fig:twiss_params}.}
\newcolumntype{Y}{>{\centering\arraybackslash}X}
\newcolumntype{M}{>{\centering\arraybackslash}m{2cm}}
\begin{center}
\begin{tabularx}{0.85\linewidth}{X Y Y}\toprule\\[-0.9em]
\textbf{Parameter} &  \textbf{$x$-plane}&  \textbf{$y$-plane}\\ \hline\\[-0.9em]
$\sigma_{x,y}$ [\SI{}{\um}]& 5.97& 6.11\\
$\beta_{x,y}$ [mm] & 4.82 & 5.41\\
$\alpha_{x,y}$ &  $0.00$&  $0.00$\\
$D_{x,y}$ [m]  &0.00 & 0.00\\
\toprule
\end{tabularx}
\end{center}
\label{table:final_params}
\end{table}

\section{Beam steering and alignment}
\label{BeamSteering}

\begin{figure}[htbp]
\centering
\includegraphics[width=0.8\linewidth]{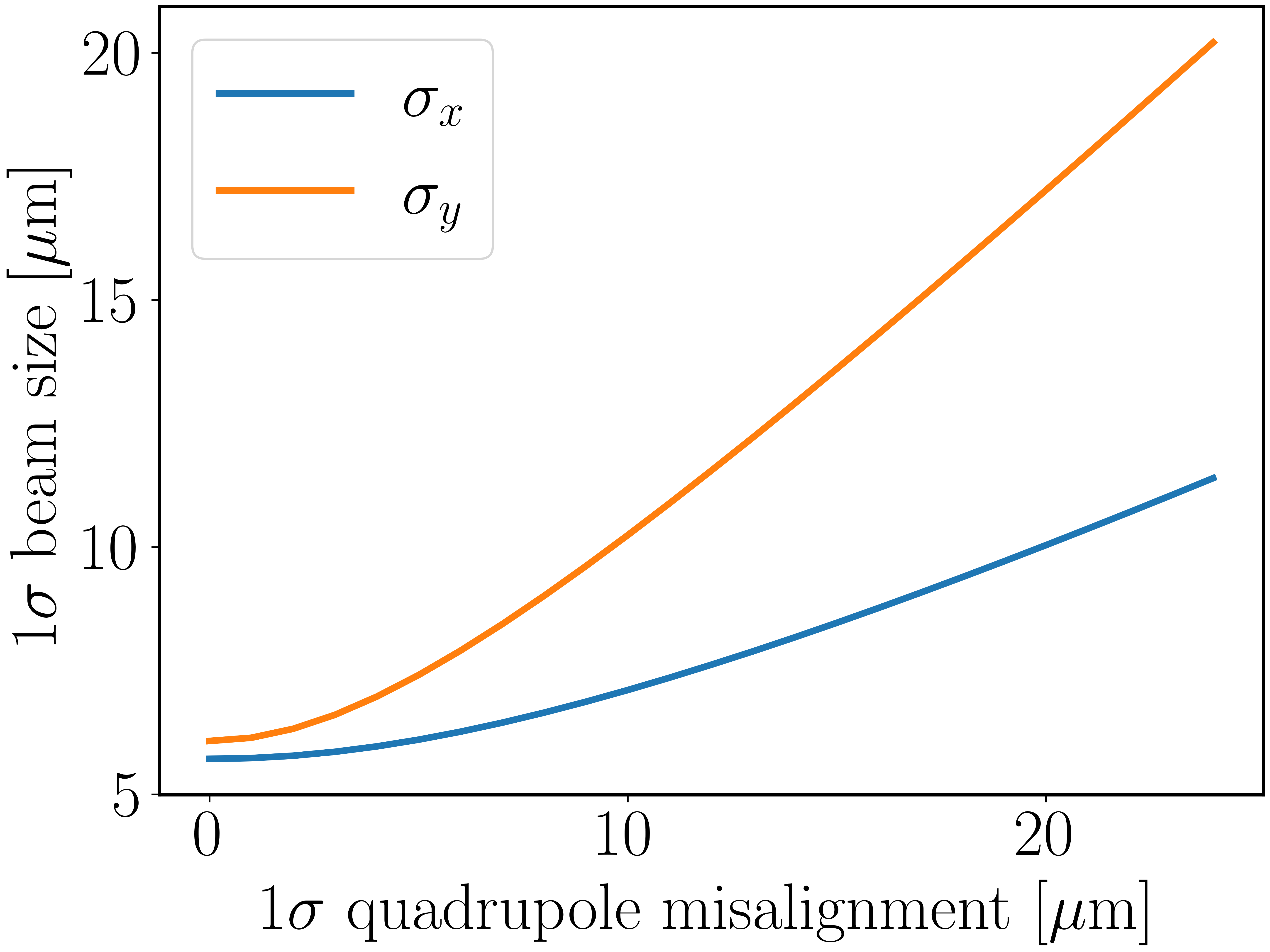}
\caption{Horizontal and vertical beam sizes averaged over 50 seeds with quadrupole misalignments sampled randomly from a Gaussian distribution with standard deviation given by the $x$-axis.}
\label{fig:quad_mis}
\end{figure}

To understand the impact of sources of error or misalignment on the beam parameters, studies of the errors individually were performed and used to specify upper bounds on the error tolerances. For example, tracking simulations as a function of the magnitude of the quadrupole misalignments are presented in Fig.~\ref{fig:quad_mis} showing that, after beam-based alignment and steering has been performed, a beam-quadrupole alignment of better than \SI{10}{\um} should be achieved. Similar studies for other error sources were used to determine suitable values to be used for error studies of the transfer line, these values are given in Table~\ref{table:errors_electron}. The magnet misalignments specified in the table are before any beam-based alignment has been performed.

\begin{figure}[htbp]
\centering
\includegraphics[width=0.99\linewidth]{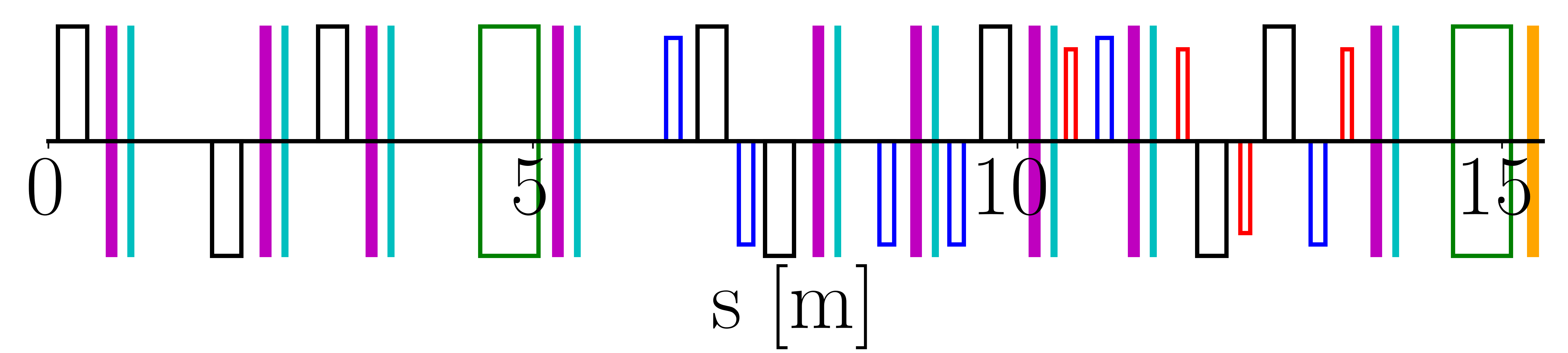}
\caption{Schematic showing the locations of BPMs (cyan), a BTV (yellow) and correctors (purple). Magnet positions are shown as outlines. The beam goes from left to right.}
\label{fig:beamsize_corr}
\end{figure}

To achieve the required levels of magnet beam alignment the magnets would need to be on movers with a step size of approximately $\SI{1}{\um}$ and with a range of 100s of microns. Multiple steering and alignment methods will be required. From simulations of the transfer line in the presence of the errors in Table~\ref{table:errors_electron} a procedure for beam-based alignment was developed. The locations of the BPMs, beam screens (BTVs) and correctors for this process are indicated in Fig.~\ref{fig:beamsize_corr}. First it is proposed to use a quad shunting technique, varying the quad strength (between 80\% and 100\%) and measuring the deflection of the beam at a downstream BPM. This deflection could be used to estimate the beam-quadrupole offset to be corrected with the magnet mover. Dispersion Free Steering (DFS) would then be used to minimise the parasitic dispersion by using the correctors to steer the beam through the centre of quadrupoles. With DFS, the beam offset is measured at all BPMs at different beam energies, the beam is then steered to both minimise the offset of the beam in the BPMs and to minimise the difference in trajectory for different beam energies. DFS is first performed with higher order magnets switched off. The deflections of the beam from sextupole or octupole offsets are likely to be difficult to resolve with the BPMs, although they will affect the beam size. Therefore, it is suggested to use the measured beam size at the injection-point BTV to quantify their alignment. We proposed to use an optimization algorithm to align the sextupoles and octupoles offsets by minimizing the beam size at the injection-point BTV. 

\begin{table}[htbp]
\caption{The r.m.s values for the distributions of errors and resolutions used for simulations of the Run 2c witness transfer line.}
\begin{center}
\newcolumntype{Y}{>{\centering\arraybackslash}X}
\newcolumntype{Z}{>{\hsize=.5\hsize}Y}
\begin{tabularx}{1\linewidth}{Y Z Z}\toprule
\textbf{Parameter} & \textbf{Error} & \textbf{Unit}\\\hline
Magnet mover position & 1 & \SI{}{\um}\\
Corrector kick & 1 & $\mu$rad\\
BPM resolution & 10 & $\mu$m \\
BTV beam size resolution & 1 & \SI{}{\um}\\
BTV position resolution & 10 & \SI{}{\um}\\
Momentum jitter & 1000 & ppm \\
Input position jitter & 10 & \SI{}{\um} \\
Dipole misalignments & 50 & \SI{}{\um} \\
Magnet field error & 10 & ppm \\
Quadrupole misalignments & 100 & $\mu$m \\
Sextupole misalignments & 100 & $\mu$m \\
Octupole misalignments & 100 & $\mu$m \\
\toprule
\end{tabularx}
\end{center}
\label{table:errors_electron}
\end{table}

The proposal for a procedure for beam steering and alignment is given below, and is estimated to be feasible in under an hour,
\begin{enumerate}
    \item quadrupole shunt – \num{2} iterations, gain \num{0.7},
    \item quadrupole shunt – \num{1} iteration, gain \num{1},
    \item DFS – higher order magnets off – \num{3} iterations, gain \num{0.7},
    \item align sextupoles and octupoles using optimization algorithm to vary magnet mover settings – \num{100} iterations.
\end{enumerate}

Multiple iterations of the quadruple-shunting and DFS methods are performed with gains set lower than unity. This is desirable to prevent the alignment process becoming unstable and also because the non-linearity of the system is sampled better with smaller steps. A lower gain is also helpful to reduce the impact of the finite BPM and BTV resolutions. For the beam-based alignment of the sextupoles and octupoles the algorithm Py-BOBYQA was employed because, as previously mentioned, it is a useful algorithm when function evaluations are costly. Here, the lifetimes of the magnet movers restrict the number of times their settings can be changed. It may also be desirable to minimise the magnitude of the steps taken by the magnet movers so that large changes are avoided when possible.  

After the setting-up procedure is performed, an efficient way to maintain the optimal conditions is to apply shot-to-shot feedback loops on trajectory using the available correctors. The change should be small enough not to significantly perturb the beam size. The same could be of course also applied to the beam size, but the main problem there is the measurements with a non-destructive technique. 

\begin{figure}[htbp]
     \centering
     \begin{subfigure}{1\linewidth}
         \centering
         \includegraphics[width=\textwidth]{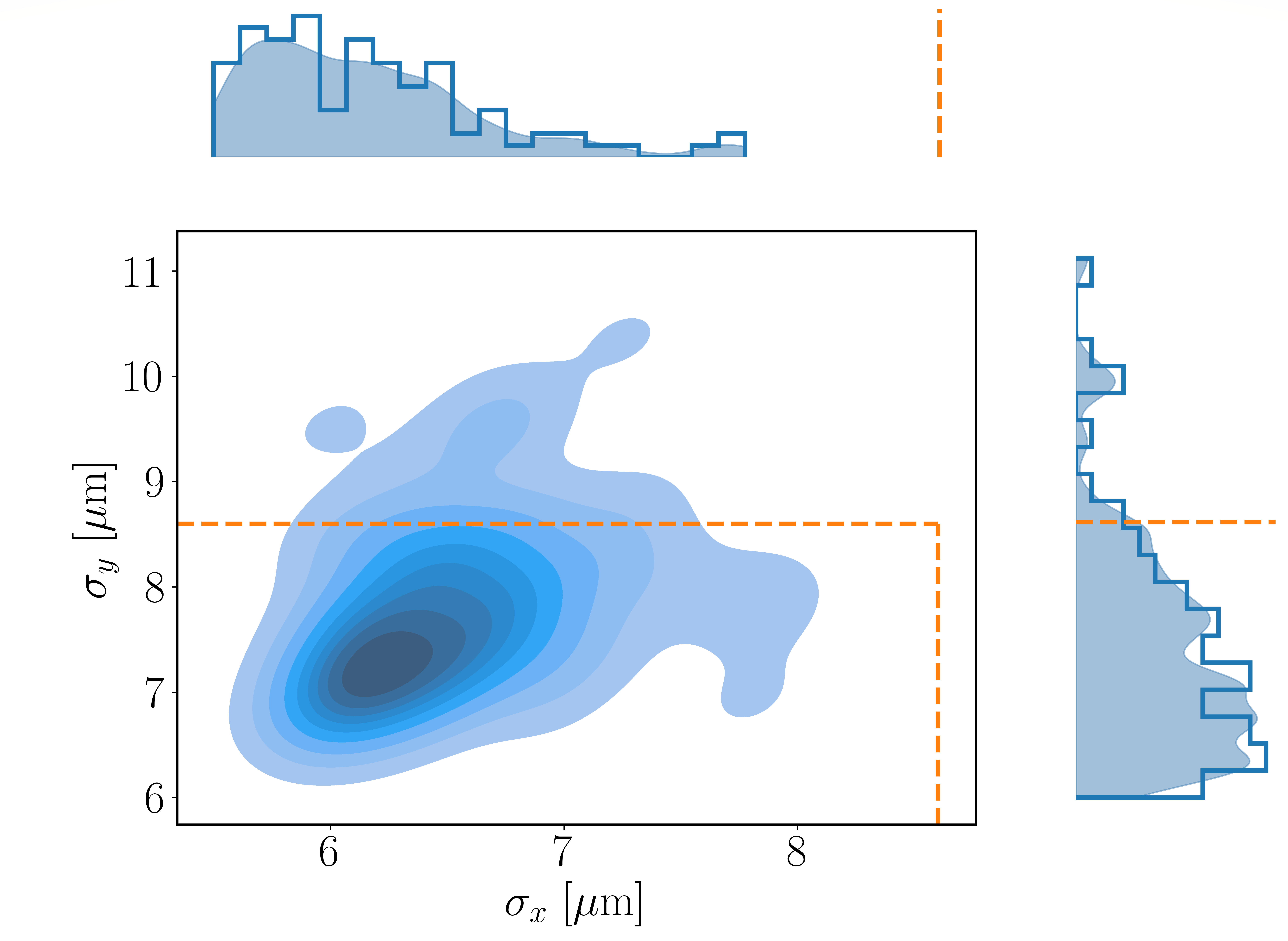}
         \caption{}
              \label{fig:heatmaps_size}
     \end{subfigure}
     \hfill
     \begin{subfigure}{1\linewidth}
         \centering
         \includegraphics[width=\linewidth]{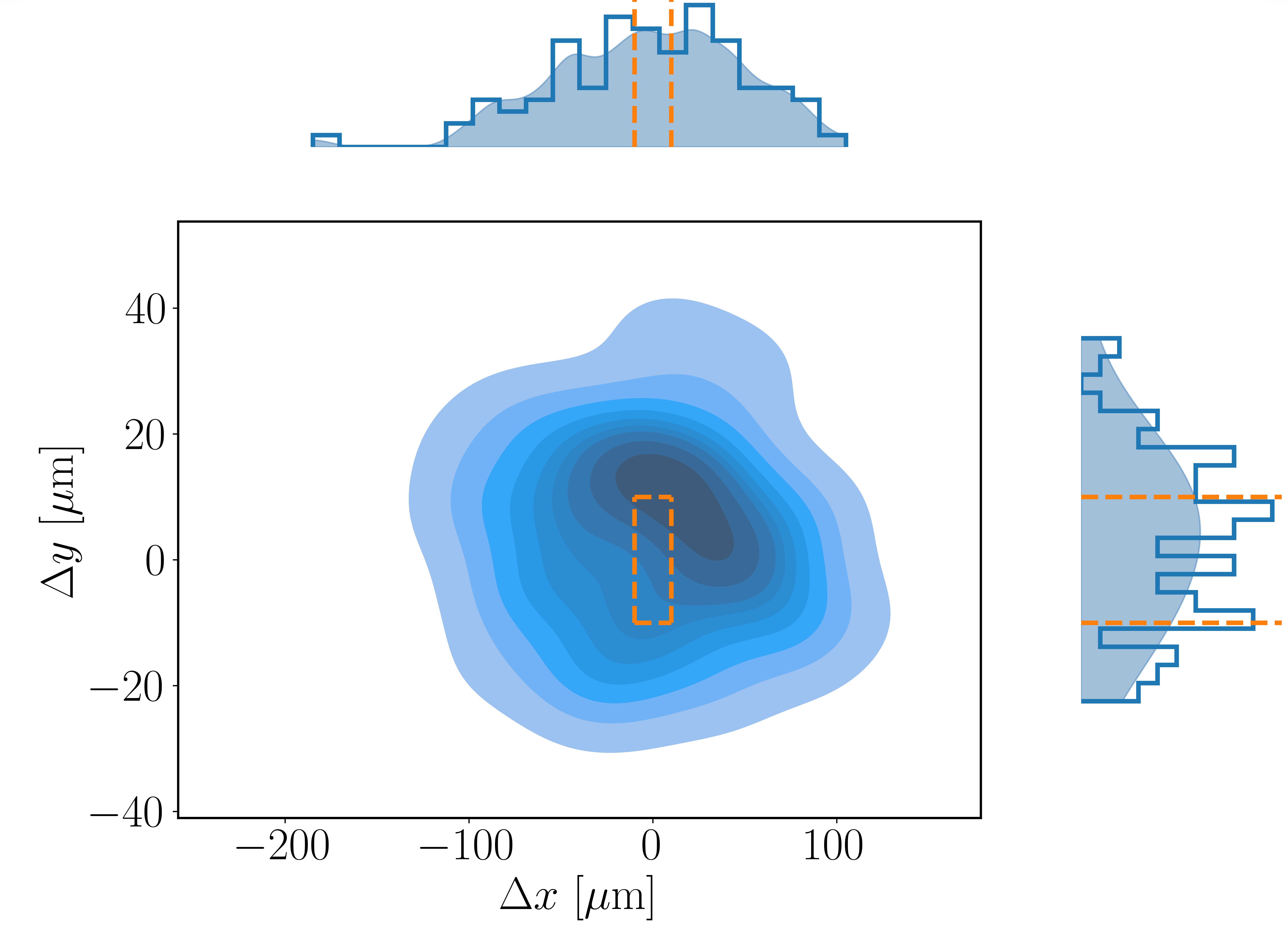}
         \caption{}
         \label{fig:heatmaps_offsets}
     \end{subfigure}
     \caption{Distributions of (a) beam sizes and (b) relative proton-electron transverse bunch offsets at the injection-point for 100 seeds with errors as defined in Table~\ref{table:errors_electron} after beam-based alignment.}

    \end{figure}

Simulations of these alignment techniques were performed with the errors and resolutions specified in Table~\ref{table:errors_electron}. As the relative proton-electron offset is of interest, also modelled was the proton beam jitter, with \SI{81.68}{\um} and \SI{10.49}{\um} r.m.s jitter in the horizontal and vertical planes, respectively. The proton beam jitter was determined from measurements of the Run 1 jitter scaled for the Run 2 configuration. 100 seeds were simulated and after the full process of beam-based alignment, 85\% of seeds satisfied the experimental beam size requirements (Fig.~\ref{fig:heatmaps_size}) and 6\% satisfied the relative offset tolerances (Fig.~\ref{fig:heatmaps_offsets}). It should be noted that the proton beam jitter was the dominant source of the offset between the beams, with the electron beam jitter and static misalignment contributing only a few microns.

\section{Scattering foils}

\begin{figure}[htbp]
\centering
\includegraphics[width=1\linewidth]{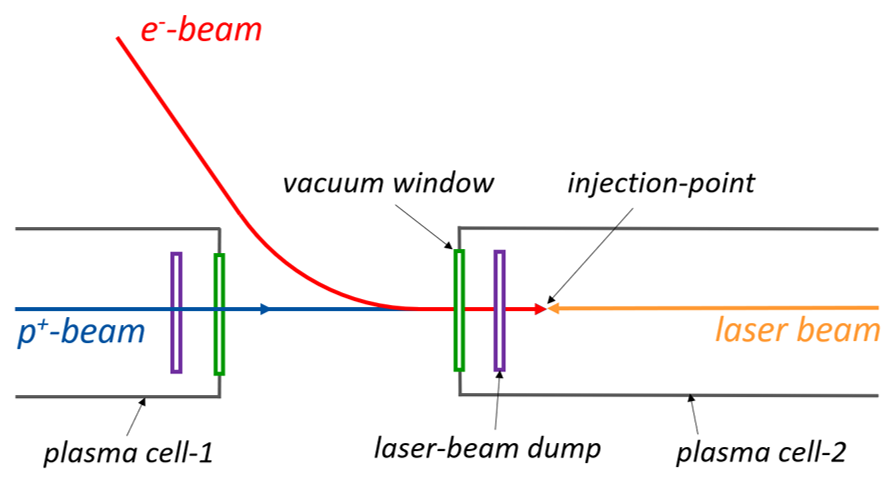}
\caption{Schematic of the witness electron injection region highlighting the locations of the vacuum window and laser-beam dump.}
\label{fig:scattering_foils}
\end{figure}

The AWAKE experiment will require two thin foils to be placed in the beamline just upstream of the focal point~\cite{verra2020study}. The first will be a vacuum window for the plasma cell and the second will be a dump for the laser used to ionize the second plasma cell, the configuration is shown in Fig.~\ref{fig:scattering_foils}. Due to the thickness and radiation length of the material traversed, and the energy of the beam, the main contributor to the change of emittance and optics is Multiple Coulomb Scattering (MCS). This translates into a net angular deflection from the original particle direction which can be modelled as Gaussian. The angular dispersion, $\theta_0$ is given by~\cite{groom2000passage}
\begin{equation}
  \theta_0 = \frac{13.6}{\beta cp}z\sqrt{\frac{x}{X_0}}[1+0.038\ln{\frac{x}{X_0}}],
  \end{equation}
where $c$ is the speed of light, $\beta=\frac{v}{c}$, $p$ is the momentum in MeV/c, $X_0$ is the material's radiation length and $x$ is the material thickness. The laser beam dump and vacuum window were modelled as two aluminium foils, with $X_0 = \SI{8.9}{\cm}$ and
\SI{1}{\mm} separation both between the two foils and between the final foil and the injection-point.

\begin{figure}[htbp]
\centering
\includegraphics[width=1\linewidth]{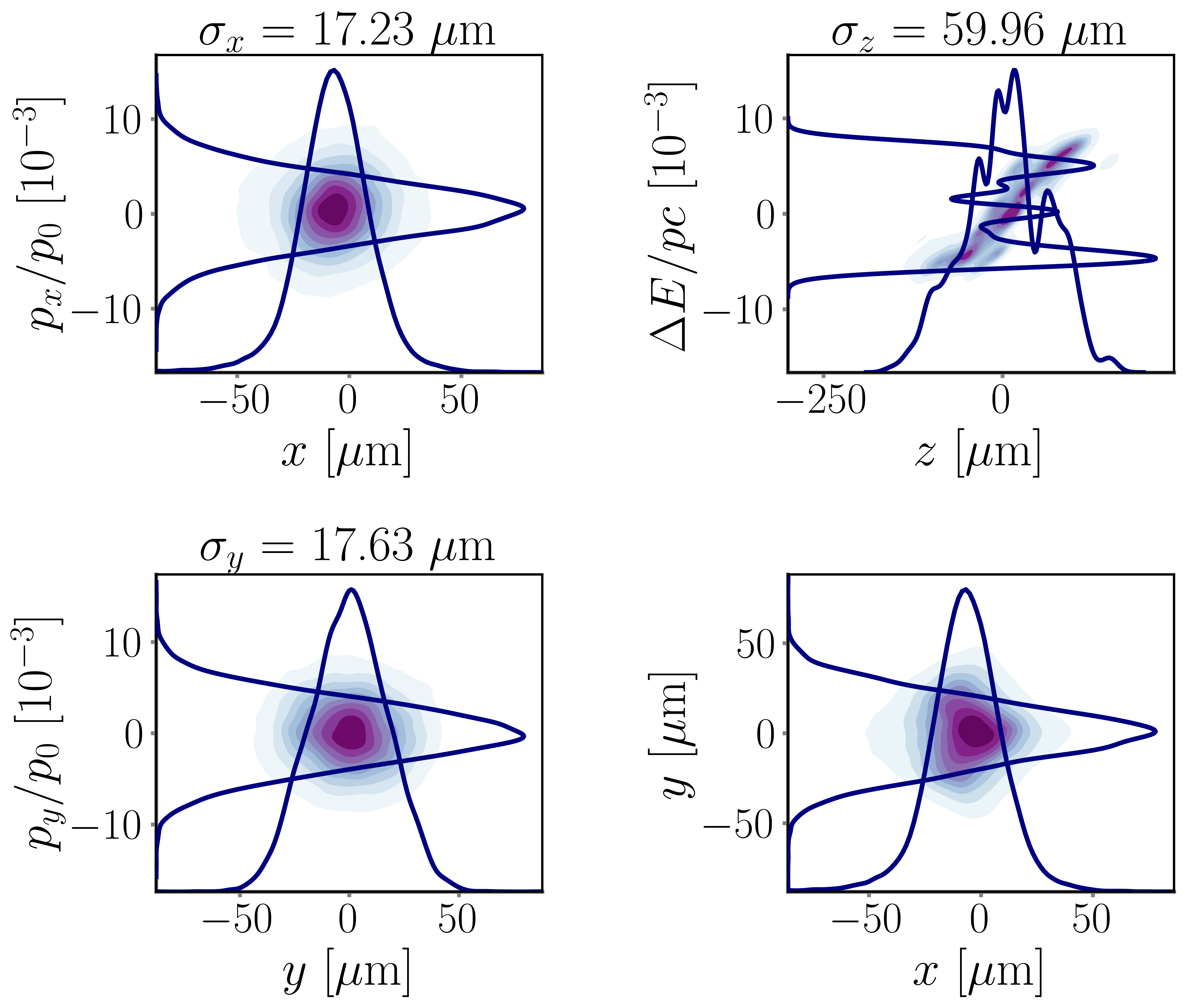}
\caption{Injection-point beam distribution from simulations of the bunch tracked through the Run 2c witness transfer line with two \SI{100}{\um} aluminium foils. Beam parameters are given in Table~\ref{table:ScatteringFoil}.}
\label{fig:beam_dist_foil}
\end{figure}

After traversing the foil the beam emittance increases, the betatron function decreases and the position of the beam waist shifts upstream. The emittance increase depends on beta function at the foil and so the the optics should be re-optimized with models of the scattering foils in the beamline so that the beam focal-point is returned to the injection-point. The Nelder-Mead algorithm was used to re-match the optics. To ensure the emittance blow-up in the $x$ and $y$ planes were equal, the term $\lvert{\sigma_x - \sigma_y}\rvert$ was incorporated in the objective function. The emittance after the scattering foil was left as a free parameter during optimization. The injection-point beam parameters for the re-matched transfer line including the scattering foils, are presented in Tab.~\ref{table:ScatteringFoil} and the beam distribution are shown in Fig.~\ref{fig:beam_dist_foil}. For these parameters, the matched beam sizes would be \SI{16.65}{\um} and \SI{16.70}{\um} horizontally and vertically respectively, which are within approximately 5\% of the achieved beam sizes.

\begin{table}[htbp]
\caption{Beam parameters at the injection-point for a transfer line with two \SI{100}{\um} aluminium foils.}
\newcolumntype{Y}{>{\centering\arraybackslash}X}
\newcolumntype{M}{>{\centering\arraybackslash}m{2cm}}
\begin{center}
\begin{tabularx}{0.85\linewidth}{X Y Y}\toprule\\[-0.9em]
\textbf{Parameter} &  \textbf{$x$-plane}&  \textbf{$y$-plane}\\ \hline\\[-0.9em]
$\sigma_{x,y}$ [\SI{}{\um}]& 17.23& 17.63\\
$\alpha_{x,y}$ &  $0.000$&  $0.000$\\
$D_{x,y}$ [m]  &0.001 & 0.000\\
$\epsilon_{x,y}$ [$\SI{}{\mm}$~mrad] & 16.96& 16.85\\
\toprule
\end{tabularx}
\end{center}
\label{table:ScatteringFoil}
\end{table}

The Nelder-Mead algorithm proved useful for making small adjustments to the transfer line optics, typically finding a solution within a few hundred iterations. It could be used, for example, to adapt the optics to produce a larger beam size, alter the beam distribution or shift the beam waist. It is foreseen that the AWAKE Run 2c experiment will scan the witness bunch parameters such as the beam size and waist position, so this capability is essential.

\section{Beam trajectory reconstruction using neural networks}
\label{NeuralNetworks}

In this section we describe how Physics-Guided Neural Networks (PGNNs~\cite{karpatne2017physics}) could be used to estimate the beam alignment for the Run 2c seeding electron line when direct measurements are not possible. Although the full design for the seeding line has not been completed, it is expected to be similar to the Run 2a electron line and consequently face similar issues. For Run 2a there were challenges with measuring the relative proton-electron alignment and so a dedicated alignment technique was developed. This could also be used for the Run 2c. 

\begin{figure}[ht]
\centering
\includegraphics[width=1\linewidth]{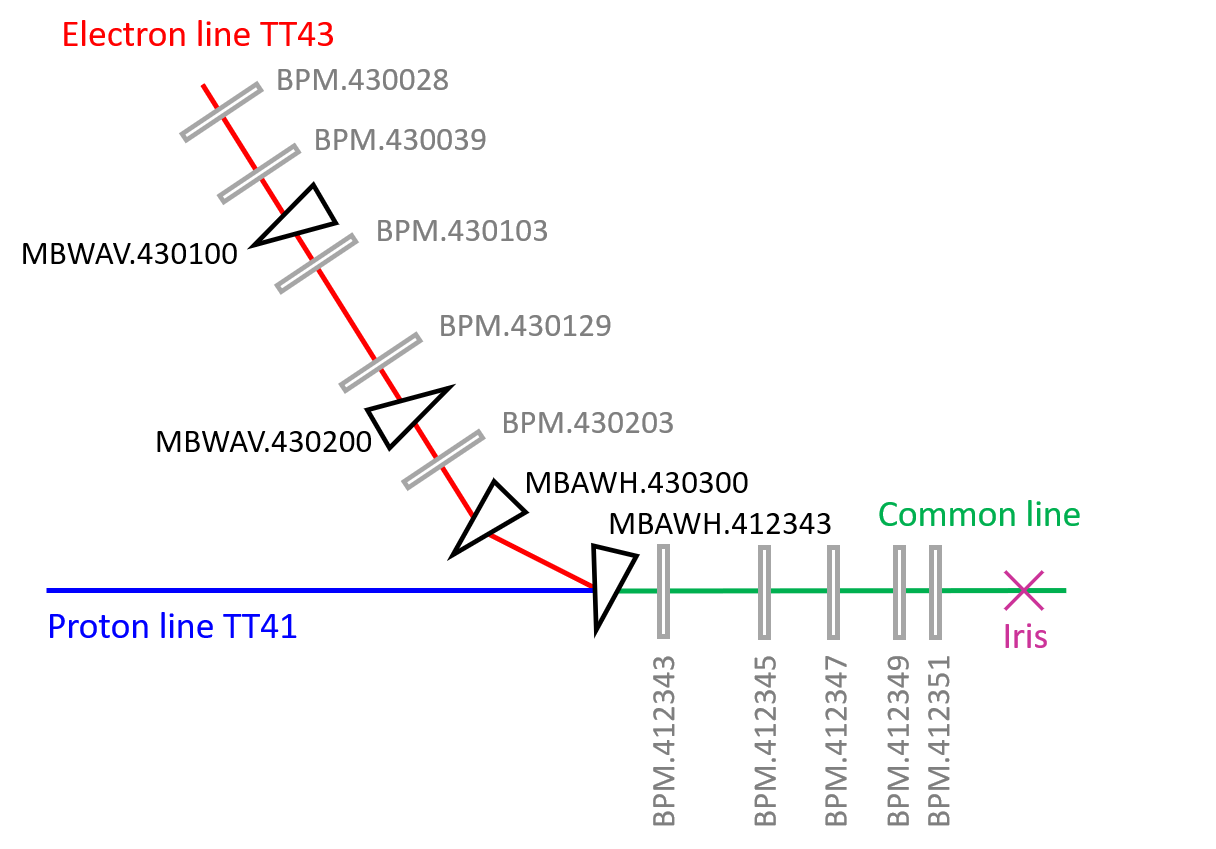}
\caption{Schematic of the proton beamline (blue), electron beamline (red) and common line (green). The BPMs relevant for this study are shown as rectangles and the electron line dipoles are given as triangles; quadrupoles are not shown in this diagram. The proton and electron beams propagate from left to right. The iris marker highlights the first iris of the plasma cell and the start of the plasma.}
\label{fig:beamlineschematic}
\end{figure}

Within the AWAKE Run 2a common line (Fig.\ref{fig:beamlineschematic}), in which the electron and proton beams co-propagate, the proton beam dominates the BPM signals so that electron measurements are not possible while there are protons. Additionally, the signals from the BPMs closest to the plasma cell are corrupted in the presence of plasma and cannot be used for either protons or electrons. Without these measurements, the beam trajectory through the common line needs to be reconstructed based on measurements from upstream in the electron line. 

The Run 2a $\sim$\SI{18}{\MeV} electron beamline (TT43) is used to inject bunches on-axis into the plasma cell. The  electron beamline has five BPMs before the common line and five BPMs within the common line (Fig.~\ref{fig:beamlineschematic}). The beam trajectory, as characterised by measurements from the first five BPMs, can be propagated into the common line using the optics model and these studies are summarised in~\cite{asmus2020predicting}. This study concluded that with this method the beam position could be predicted at the final BPM, BPM.412351, with r.m.s errors of $\sim$\SI{370}{\um} horizontally and $\sim$\SI{150}{\um} vertically. Here we describe a method to improve these predictions with the addition of PGNNs.

\subsection{Physics-Guided Neural Network}

PGNNs use predictions from a physics-model of the system alongside measured data as the inputs to the NN. In cases where the measured data is inaccurate or noisy, the physics model may give better prediction, whereas, if the model has significant shortcomings in representing the system, the measured data would be more trustworthy. Typically, the situation would be somewhere between these two cases and by combining the model with the measured data one can get benefits from them both.

\subsection{Beam trajectory predictions}

\begin{figure}[htbp]
     \centering
     \begin{subfigure}{0.45\textwidth}
         \centering
         \includegraphics[width=\textwidth]{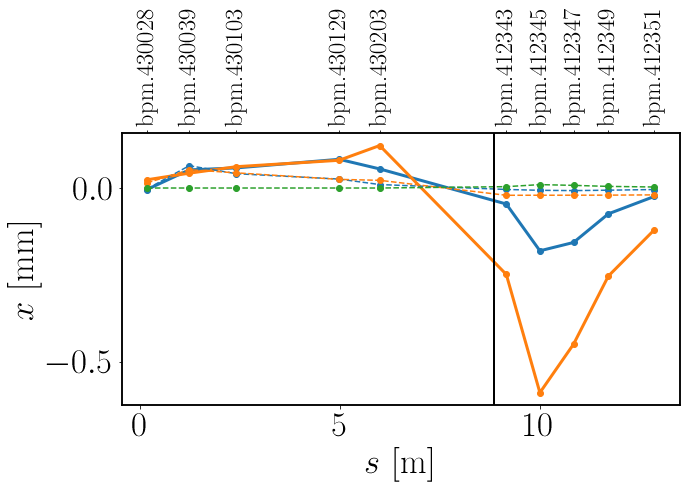}
         \caption{}
     \end{subfigure}
     \hfill
     \begin{subfigure}{0.48\textwidth}
         \centering
         \includegraphics[width=\textwidth]{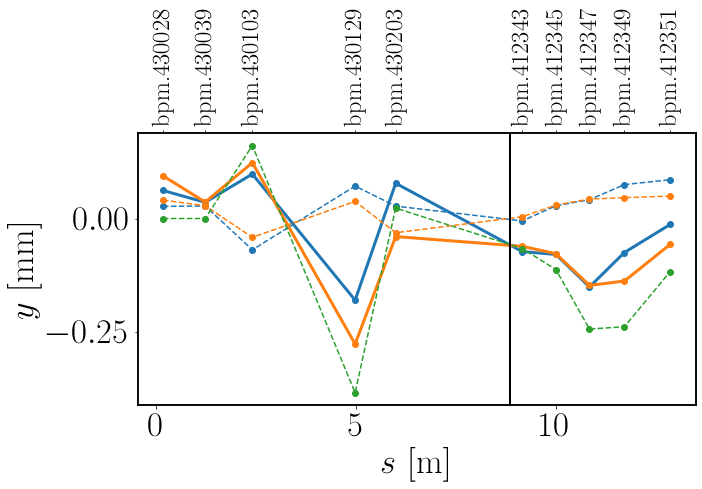}
         \caption{}
     \end{subfigure}

        \caption{Measured horizontal (a) and vertical (b) beam position offsets with respect to the mean trajectory for a single event. Measured values are shown as blue lines and predicted values as orange lines, where the bold lines show beam positions and the dashed lines show the betatron component. The dispersion components are shown as dashed green lines. The start of the common line is denoted by a vertical black line.}
        \label{fig:single_optics}
\end{figure}

The beam trajectory through the common line can be predicted for every pulse by propagating the electron trajectory using the transfer line model with the following method. Firstly, with the plasma off and no proton beam, the mean beam trajectory was measured at all ten BPMs in the electron line. The mean trajectory was subtracted from the measurements so as to keep the pulse-to-pulse variation only. Secondly, the momentum offset, $\delta_p$, was calculated for every pulse, using the method described in~\cite{chiara2019momentum}, by exploiting that the BPMs, BPM.30103 (103) and BPM.430129 (129), have a phase advance of almost $\pi$ between them so that,
\begin{equation}
    \delta_p \simeq \frac{\sqrt{\beta_{103}}y_{129} + \sqrt{\beta_{129}}y_{103}}{\sqrt{\beta_{103}}D_{129} + \sqrt{\beta_{129}}D_{103}}.
\end{equation}
From $\delta_p$, the dispersion contribution was calculated and subtracted from the beam trajectory leaving the betatron contribution. The optics model was used to propagate the betatron contribution into the common line and, finally, the dispersion component and mean trajectories were added back to get beam position estimates. The predicted beam trajectories for a representative pulse are shown in Fig.~\ref{fig:single_optics}. The discrepancy between the horizontal trajectory measurements and predictions, seen in Fig.~\ref{fig:single_optics}, were due to a difference between the optics model and the beamline, where in~\cite{asmus2020predicting} this is hypothesised to be an offset of quadrupole 430311.

A PGNN was tested to predict the residual errors from the optics model propagation of the beam trajectory. The PGNN had thirty features consisting of the first five BPM measurements, both horizontally and vertically, and the corresponding beam trajectory predictions for all ten BPMs. The input values
\begin{equation}
\begin{split}
  [x_{1}^\mathrm{meas.},..., x_{5}^\mathrm{meas.},x_{1}^\mathrm{pred.},...,x_{10}^\mathrm{pred.}, \\ y_{1}^\mathrm{meas.},...,y_{5}^\mathrm{meas.},y_{1}^\mathrm{pred.},...,y_{10}^\mathrm{pred.}].
  \end{split}
\end{equation}
were normalized to lie in the range 0 to 1. The NN output comprised the errors on the optics model predictions compared with the BPM measurements for the final five BPMs,
\begin{equation}
\begin{split}
 & [x_{6}^\mathrm{meas.}-x_{6}^\mathrm{pred.},..., x_{10}^\mathrm{meas.}-x_{10}^\mathrm{pred.},\\ &y_{6}^\mathrm{meas.}-y_{6}^\mathrm{pred.},..., y_{10}^\mathrm{meas.}-y_{10}^\mathrm{pred.}].
  \end{split}
\end{equation}
As the PGNN output is compared with BPM measurements to calculate the PGNN output error, the resolution of the BPMs sets a limit to the PGNN performance that can be measured. To propagate the predictions from the BPMs to the iris, the beam angle was calculated from the ballistic trajectory through the final two BPMs. This propagation should take into account the effect on the beam trajectory from the Earth's magnetic field.

The PGNN had six hidden layers with the number of nodes per layer stepping from \num{30} to \num{60} and back to \num{30} in steps of \num{10}. The hidden layers had $\tanh$ activation functions. During training the learning rates were decreased step-wise throughout training and this process was optimized empirically. The data were split 80\%/20\% split into training and test data, with a further 10\% of the training data used for validation to highlight any over-fitting of the model. A batch size of 64 was used for training and of order \num{1000} epochs. An MSE loss was used to quantify the PGNN performance.

\subsection{Results}

\begin{figure}[htbp]
\centering
\includegraphics[width=1\linewidth]{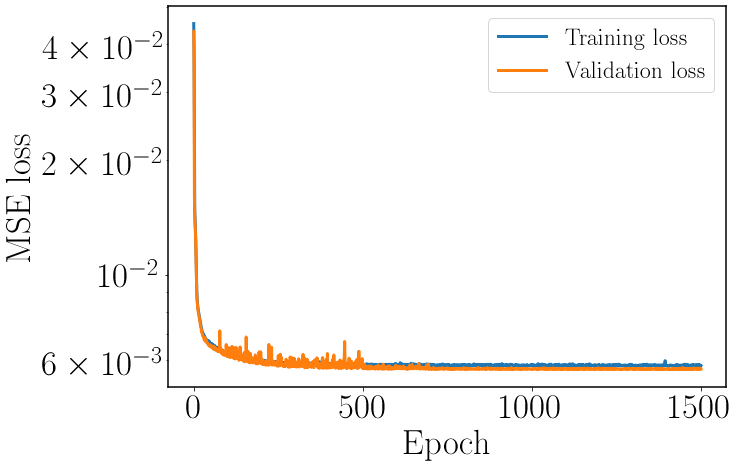}
\caption{Training and validation MSE loss vs. epoch.}
\label{fig:pgnn_train}
\end{figure}

PGNNs were tested on data at three different charges, \SI{300}{pC}, \SI{650}{pC} and \SI{750}{pC}, as measured with a Farady cup. The BPM resolution scales with the BPM signal-to-noise ratio and, consequently, with the beam charge so that these data could be used to study the variation in PGNN performance with BPM resolution. Three separate NNs were trained for the three charges. The \SI{750}{pC} data set had $\sim$4000 events with 80\% used for training and validation. The training and validation MSE losses are shown in Fig.~\ref{fig:pgnn_train} for 1500 epochs of training.

\begin{figure}[htbp]
     \centering
     \begin{subfigure}{0.4\textwidth}
         \centering
         \includegraphics[width=\textwidth]{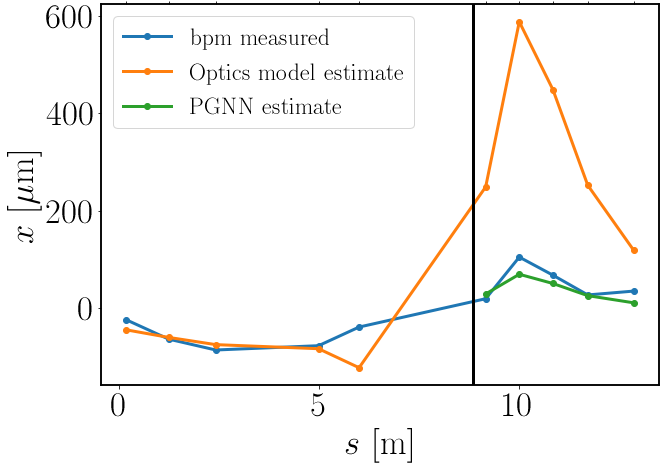}
         \caption{}
     \end{subfigure}
     \begin{subfigure}{0.42\textwidth}
         \centering
         \includegraphics[width=\textwidth]{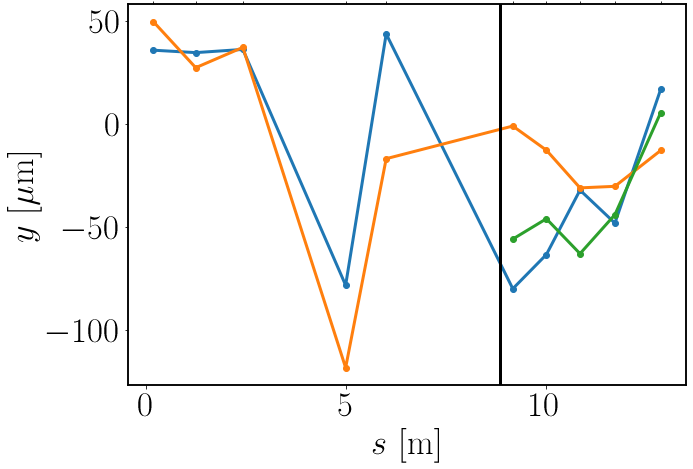}
         \caption{}
     \end{subfigure}
        \caption{A comparison of the optics model beam trajectory prediction (orange) and the PGNN prediction (green) with the measured high-charge BPM data (blue), shown for the horizontal (a) and vertical (b) planes.}
        \label{fig:pgnn_single_high}
\end{figure}

The predictions of the PGNN compared with the measured data and optics model prediction are given in Fig.~\ref{fig:pgnn_single_high} for a single test event. The PGNN performances are presented in Tables~\ref{tab:high_charge_horizontal} and~\ref{tab:high_charge_vertical} for the three charges, where the degradation in measured PGNN performance with decreasing charge can be seen. The AWAKE BPMs are expected to have resolutions of up to \SI{20}{\um} which agree well with the high-charge results.

For the low charge data the BPM signal-level was approximately a tenth of that for the high charge data. With this lower resolution, neither the optics model prediction nor the PGNN prediction are measured to perform much better than the level of the beam jitter. This is likely as a result of the BPM resolution being of the same order as the measured beam jitter, meaning that the jitter measurements are resolution-limited. In this case, better results may be achieved by assuming always the mean beam trajectory rather than trying to predict the trajectory pulse-to-pulse.

\begin{table*}[htbp]
\caption{Horizontal: Measured jitter for three charge settings compared with the r.m.s error from the optics model predictions and the r.m.s error from the PGNN predictions. The low-charge measurements show evidence of being resolution-limited.}
\centering
\newcolumntype{Y}{>{\centering\arraybackslash}X}
\newcolumntype{Z}{>{\centering\arraybackslash\hsize=0.4
\hsize}X}
\begin{tabular*}{\linewidth}{@{\extracolsep{12.5pt}}p{0.07\textwidth}>{\centering\arraybackslash}p{0.07\textwidth}>{\centering\arraybackslash}p{0.07\textwidth}>{\centering\arraybackslash}p{0.07\textwidth}>{\centering\arraybackslash}p{0.07\textwidth}>{\centering\arraybackslash}p{0.07\textwidth}>{\centering\arraybackslash}p{0.07\textwidth}>{\centering\arraybackslash}p{0.07\textwidth}>{\centering\arraybackslash}p{0.07\textwidth}>{\centering\arraybackslash}p{0.07\textwidth}}\toprule

\textbf{BPM} & \multicolumn{3}{c}{\textbf{Position jitter [\SI{}{\um}]}} & \multicolumn{3}{c}{\textbf{Optics model r.m.s error [\SI{}{\um}]}} & \multicolumn{3}{c}{\textbf{PGNN r.m.s error [\SI{}{\um}]}}\\

& low&medium&high&low&medium&high &low&medium&high \\

\cline{2-4} \cline{5-7} \cline{8-10}  

412343 &  147.6&63.6&509.5 & 310.0&195.6&136.6 & 144.4 &26.6&25.9 \\ 

412345 &  155.8&143.7&1201.1 &  673.5&457.0&324.4 &  148.0&53.4&53.8 \\

412347 &  165.5&99.9&748.1 & 578.3&343.5& 128.1 & 164.2 &41.2& 41.8\\ 
412349 &  141.1&66.0&416.9 & 438.9 &195.8& 71.4 & 142.5 &31.4&25.3 \\ 

412351 &  131.6&52.5&156.1 & 386.7 &98.0&55.2 & 133.9&28.8& 19.4 \\

\toprule
\end{tabular*}
\label{tab:high_charge_horizontal}
\end{table*}

\begin{table*}[htbp]
\caption{Vertical: Measured jitter for three charge settings compared with the r.m.s error from the optics model predictions and the r.m.s error from the PGNN predictions. The low-charge measurements show evidence of being resolution-limited.}
\centering
\newcolumntype{Y}{>{\centering\arraybackslash}X}
\newcolumntype{Z}{>{\centering\arraybackslash\hsize=0.4
\hsize}X}
\begin{tabular*}{\linewidth}{@{\extracolsep{12.5pt}}p{0.07\textwidth}>{\centering\arraybackslash}p{0.07\textwidth}>{\centering\arraybackslash}p{0.07\textwidth}>{\centering\arraybackslash}p{0.07\textwidth}>{\centering\arraybackslash}p{0.07\textwidth}>{\centering\arraybackslash}p{0.07\textwidth}>{\centering\arraybackslash}p{0.07\textwidth}>{\centering\arraybackslash}p{0.07\textwidth}>{\centering\arraybackslash}p{0.07\textwidth}>{\centering\arraybackslash}p{0.07\textwidth}}\toprule

\textbf{BPM} & \multicolumn{3}{c}{\textbf{Position jitter [\SI{}{\um}]}} & \multicolumn{3}{c}{\textbf{Optics model r.m.s error [\SI{}{\um}]}} & \multicolumn{3}{c}{\textbf{PGNN r.m.s error [\SI{}{\um}]}}\\

& low&medium&high&low&medium&high &low&medium&high \\

\cline{2-4} \cline{5-7} \cline{8-10}  

412343 & 175.5&97.5&73.5 & 147.2&32.9&69.6  & 136.7&19.7&15.1  \\ 
412345 & 184.6&94.6&28.4   & 148.2&25.0&46.4  & 143.4&19.1&15.4 \\
412347 & 247.4&165.5&142.9  & 189.3&39.5&89.4  &  165.4&22.5&21.1\\ 
412349 & 224.8 &131.9&242.3 & 188.2 &57.6&95.2 & 152.1&21.1&18.7 \\ 
412351 & 233.9&42.9& 148.0 & 180.2&56.4&62.3 & 143.7 &19.6&16.6\\

\toprule
\end{tabular*}

\label{tab:high_charge_vertical}
\end{table*}

The results from propagating the beam trajectory to the iris are presented in Table~\ref{tab:propagation}. There is a clear improvement in the position and angle predictions at medium and high charge with the PGNN. The resolution of the position measurement extrapolated to the iris can be calculated from the geometry of the system. This can then be compared with the measured PGNN performance. For example, if the final two BPMs have a resolution of \SI{20}{\um} at high charge, this would correspond to a \SI{48}{\um} resolution at the iris. The PGNN error propagated to the iris is in good agreement with this. If we assume that the low-charge results are resolution-limited, then the final two BPMs have a resolution $\sim$\SI{140}{\um}, which would correspond to a \SI{340}{\um} resolution at the iris.

\begin{table*}[hbtp]
\caption{Angle jitter and position jitter propagated to the iris, both calculated using measurements at the last two BPMs. Comparison with the r.m.s errors from the optics model predictions and PGNN model predictions.}
\centering
\begin{tabular*}{\linewidth}{@{\extracolsep{12.5pt}}p{0.3\textwidth}>{\centering\arraybackslash}p{0.08\textwidth}>{\centering\arraybackslash}p{0.08\textwidth}>{\centering\arraybackslash}p{0.08\textwidth}>{\centering\arraybackslash}p{0.08\textwidth}>{\centering\arraybackslash}p{0.08\textwidth}>{\centering\arraybackslash}p{0.08\textwidth}}\toprule
 & \multicolumn{3}{c}{\textbf{Horizontal}} & \multicolumn{3}{c}{\textbf{Vertical}}\\
& low&medium&high&low&medium&high\\
\cline{2-4} \cline{5-7}
Angle jitter [$\mu$rad]&    200.1  &117.6&118.7 &      180.5  &90.9&113.2\\
Angle optics error (r.m.s) [$\mu$rad]&    170.2 &92.1&60.9 & 178.7&25.2&45.5 \\
Angle PGNN error (r.m.s) [$\mu$rad]&   163.9&24.0& 21.8 &    178.3 &23.8&19.1\\\hline
Iris jitter [$\mu$m] & 603.9&161.3& 488.0 &  494.8&181.9&194.2\\
Iris optics error (r.m.s) [$\mu$m] & 526.7&133.1&161.04 & 507.9&86.1&94.8\\
Iris PGNN error (r.m.s) [$\mu$m]  & 445.8&68.9&57.6  & 495.9&66.0&52.8\\\hline
\toprule
\end{tabular*}

\label{tab:propagation}
\end{table*}

\begin{figure}[htbp]
\centering
\includegraphics[width=0.9\linewidth]{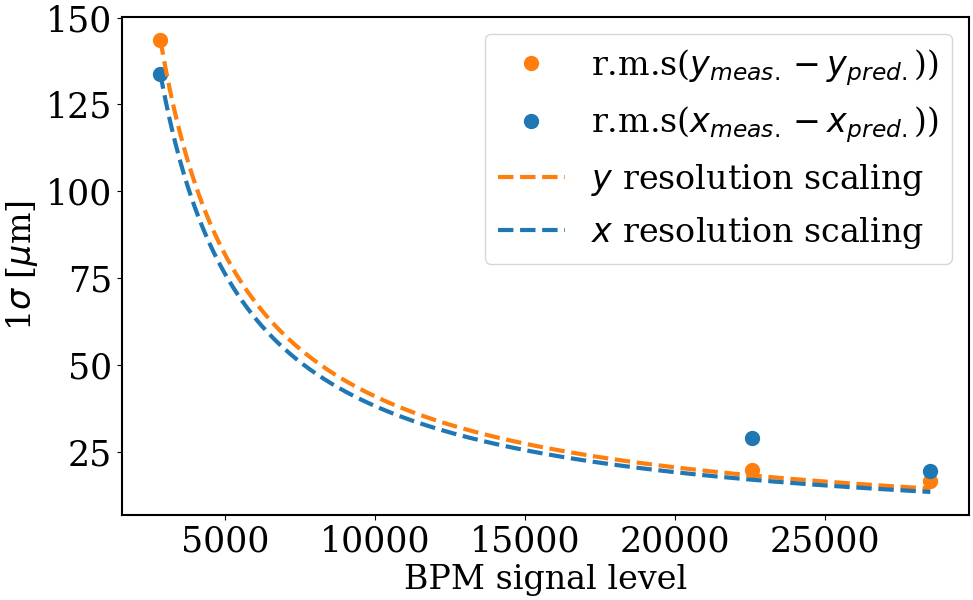}
    \caption{The data points show the horizontal and vertical r.m.s error between the measured and prediction positions at BPM 51; the resolution scaling from low charge is given as a dashed line under the assumption that the low-charge results were resolution-limited.}
      \label{fig:res_scale}
\end{figure}

The resolution is expected to scale linearly with the BPM signal-to-noise ratio, where the sums of the BPM signals for the three charges were \num{2850} (\SI{300}{pC}), \num{22500} (\SI{650}{pC}), \num{28500} (\SI{750}{pC}). The best PGNN performances, as measured at BPM 51, are presented in Fig.~\ref{fig:res_scale} along with the estimated resolution scaling calculated by assuming the performance of the low-charge case was resolution-limited. There is good agreement vertically but horizontally the higher charge predictions under-perform compared to the resolution scaling. The BPMs with the largest horizontal jitters also demonstrate the poorest horizontal predictions, requiring further study.

\section{Generalization of the techniques developed}

The usage of genetic algorithms and numerical optimizers for beam line design is relatively new and still not fully exploited. In cases where considering only linear optics is not sufficient and accounting for all aberrations is non-trivial, numerical optimizers can aid in finding a working solution. Supervision is clearly still needed and any additional information, included as constraints, is invaluable. Establishing constraints, often with a physical basis, can help converge towards more elegant solutions which are more robust to errors. 

Potential developments of this technique could be the employment of hierarchical structures in multi-layer optimizers to further automatize the design of transfer lines.

It has been shown that PGNNs can be used to reconstruct the beam trajectory through the Run 2a AWAKE common line. This method could be adapted for use with the Run 2c seeding electron line and even developed into an application to give real-time predictions of the relative alignment between the proton and electron beams. Ideally, the Run 2c seeding electron line would also have BPMs in the dog-leg at $\pi$-phase-advance so that the momentum offset could again be easily measured. 

The help of PGNNs for trajectory predictions is a very general concept for accelerators. As optics knowledge is very frequently available, this could represent a way to circumvent invasive or expensive beam diagnostics.  A natural way to extend this further would be to exploit raw BPM waveforms to try to estimate the beam size.

\section{Conclusion}

In this paper we have described how numerical optimization and neural networks were used during the design of the AWAKE Run 2c electron transfer lines.

The baseline design of a \SI{150}{\MeV} electron transfer line to inject witness bunches into the second plasma cell for AWAKE Run 2c was presented. The spatial constraints and experimental requirements for micron-level beam size and stability were challenging and various optimization techniques were used during the design process. Genetic algorithms were exploited to produce an initial dog-leg design and various local optimization algorithms were employed to advance this design. The strong focusing that was required led to the rise of significant non-linerarities, such as betatronic chromatic effects and detuning with amplitude. Sextupoles and octupoles were added to mitigate these effects and their positions and strengths were optimized with the Powell algorithm. 

The effects of errors and misalignments on the beam size and stability were studied and a correction process was developed and simulated. These studies suggest that after beam-based alignment, 85\% of the pulses should satisfy the experimental beam size specification. For the relative proton-electron beam misalignment, only 4\% of pulses were within the tolerances, but this was dominated by the proton beam jitter. Optimization algorithm Py-BOBYQA was studied for the beam-based alignment of the sextupoles and octupoles using beam size measurements from a BTV at the injection-point. While this was successful in simulation, further studies into the feasibility of this method are required, in particular, accounting for the mover limitations and lifetime. 

An optimization framework based on the Nelder-Mead algorithm was created and this facilitated the re-optimization of the transfer line optics if minor adjustments were needed. This was used to re-match the transfer line to include two thin scattering foils upstream of the focal point. These foils increased the emittance so that the matched beam size was of order $\sim$\SI{17}{\um}.

Studies were also presented towards developing a method for predicting the relative proton-electron alignment for the Run 2c seeding line. For this transfer line, the beam trajectory would need reconstructing through regions with no available direct position measurements. A method suitable for Run 2c was tested on the Run 2a beamline and the addition of PGNNs was shown to offer significant improvements compared with using only the optics model. The results at high charge were consistent with the expected resolution-limit of the BPMs, with PGNN r.m.s errors at the final BPM of $<\SI{20}{\um}$. The performance of the vertical predictions at different charges scaled well with the expected resolution, however, the horizontal results deviated, thus requiring further study.

\section{Acknowledgements}

We would like to thank G.~Zevi Della Porta and L.~Verra for their assistance in providing the beam time and conditions to take these data. We are grateful also to J.~Farmer, M.~Weidl, P.~Muggli for many helpful discussions: your expertise were greatly appreciated. Additionally, we would like to thank E.~Gschwendtner for her guidance of the project. 

Finally, we are thankful for the financial support of this research from the Science and Technology Facilities Council (AWAKE-UK, Cockcroft Institute
core, John Adams Institute core, and UCL
consolidated grants), United Kingdom. 


\providecommand{\noopsort}[1]{}\providecommand{\singleletter}[1]{#1}%

\end{document}